\documentclass[reprint,amsmath,amssymb,
prb,floatfix,aps]{revtex4-2}
\usepackage{graphicx}% Include figure files
\usepackage[dvipsnames]{xcolor}
\usepackage{comment}
\usepackage{dcolumn}% Align table columns on decimal point
\usepackage{bm}% bold math
\usepackage{physics}
\usepackage[normalem]{ulem}
\newcommand{\orcid}[1]{\href{https://orcid.org/#1}{\includegraphics[width=8pt]{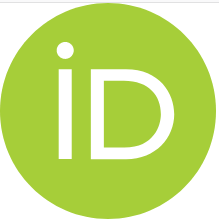}}}
\usepackage[colorlinks,urlcolor=goodgreen,citecolor=blue,linkcolor=goodred]{hyperref}
\definecolor{orange}{rgb}{1,0.5,0}
\definecolor{goodgreen}{rgb}{0.1,0.5,0}
\definecolor{goodred}{rgb}{0.7,0,0}

\let\oldepsilon\epsilon \let\epsilon\varepsilon \let\varepsilon\oldepsilon

\renewcommand\vec{\boldsymbol}

\newcommand{\SB}[1]{{\color{magenta}[SB: #1]}}

\newcommand{\TK}[1]{{\color{orange}[TK: #1]}}

\begin{document}
\title{General boundary conditions for the quasiclassical kinetic equations in the presence of magnetism and unconventional superconductivity}
\author{Tim Kokkeler \orcid{0000-0001-8681-3376}}
\email{tim.h.kokkeler@jyu.fi}
\affiliation{Department of Physics and Nanoscience Center, University of Jyväskylä, P.O. Box 35 (YFL), FI-40014 University of Jyväskylä, Finland}
\author{Yukio Tanaka\orcid{0000-0003-1537-4788}}
\affiliation{Department of Applied Physics, Nagoya University, 464-8603 Nagoya, Japan\\
}
\affiliation{Research Center for Crystalline Materials Engineering, Nagoya University, 464-8603 Nagoya Japan}

%\author{Tero T. Heikkilä \orcid{0000-0002-7732-691X}}
%\affiliation{Department of Physics and Nanoscience Center, University of Jyväskylä, P.O. Box 35 (YFL), FI-40014 University of Jyväskylä, Finland}
\author{F. Sebastian Bergeret\orcid{0000-0001-6007-4878}}
\email{fs.bergeret@csic.es}

\affiliation{Centro de Física de Materiales (CFM-MPC) Centro Mixto CSIC-UPV/EHU,
E-20018 Donostia-San Sebastián, Spain}
\affiliation{Donostia International Physics Center (DIPC), 20018 Donostia--San
Sebastián, Spain}

\begin{abstract}
    We present  boundary conditions for the quasiclassical Green's functions describing  spin-activeinterfaces between two materials, at least one of which is in the diffusive regime. These boundary conditions are developed with the help of the action formalism, and provide clean and easily tractable expressions. We connect our expressions to previous literature in limiting cases and describe normal transport, magneto-resistance transport and spin-precession through the boundary. By using the Keldysh-Nambu-spin formalism, our equations are valid both in the normal and superconducting state. The construction through the action formalism also allows us to derive the corresponding full counting statistics for charge currents, thermal currents and spin currents. We use this boundary condition to consider the anomalous proximity effect, and to study the Josephson coupling between triplet and spin-split superconductors through a magnetic interface. We find an analytic expression for the zero bias conductance in junctions with a \textit{p}-wave superconductor and a spin-filter, and show how it depends on the relative orientation of the exchange field and \textit{d}-vector. For Josephson junctions, we show that, in agreement with time-reversal symmetry, the conditions on the relative orientations of the spins in the materials for the appearance of anomalous Josephson currents, are distinctly different for two spin-split superconductors and a spin-split superconductor with a triplet superconductor. Moreover, we show that the anomalous Josephson currents are only accompanied by a diode effect if both transparencies for spin up and down are nonzero and distinct.
\end{abstract}

\maketitle
\onecolumngrid\
\twocolumngrid\
\section{Introduction}
%The disappearance of resistance below a the critical temperature is one of the most important features of superconductors, it makes them promising materials for the new generation of electronic devices. In such devices it is inevitable to combine superconductors with other materials such as normal metals or ferromagnet. This interplay 
%Transport in junctions between two superconductors or a superconductor and a non-superconducting material is one of the most important features of superconducting electronics. Because many  materials that are developed are not clean and hence have a large scattering rate, there have been efforts to describe such junctions within the quasiclassical Usadel formalism.

%Recently, a phenomenological approach to deriving kinetic equations has been introduced, valid from zero temperature, where metals may be deep in the superconducting state, up to arbitrarily high temperatures, where transport is governed by non-equilibrium effects. This method is based on the insight that the effective Keldysh action for soft modes in a disordered system can be constructed directly from symmetry considerations.

Transport in mesoscopic metals and superconductors is most conveniently described within the quasiclassical formalism. Specifically, for materials in which transport is diffusive, the celebrated Usadel equation \cite{usadel1970generalized} is widely employed.
%In recent times, 
The Usadel equation can be  derived from an action formalism, via the Keldysh nonlinear sigma model action, an effective action for the soft diffusion modes of the material \cite{kamenev2009keldysh,feigel2000keldysh,kamenev2023field}. The saddle point condition of the nonlinear sigma model action gives the quasiclassical equations. %In this way, the Usadel equation has been extended beyond the lowest order in quasiclassics \cite{virtanen2021magnetoelectric,virtanen2022nonlinear,kokkeler2024nonreciprocal,kokkeler2025quantum}.
%The resulting kinetic equation, known as the generalized Usadel equation, extends the well-known quasiclassical framework to incorporate not only superconductivity but also spin-orbit coupling and forms of unconventional magnetism, such as altermagnetic order.

Once the quasiclassical equations are obtained, describing hybrid interfaces between materials with distinct ground states requires appropriate boundary conditions for these equations. For example, in the low-transparency limit, the Kupriyanov–Lukichev boundary conditions are widely used \cite{kuprianov1988influence}. A generalization of these conditions to arbitrary interface transparency was presented by Nazarov \cite{nazarov1999novel} within the framework of circuit theory for superconducting hybrid structures.
These boundary conditions have been further generalized to describe interfaces with unconventional superconductors \cite{tanaka2003circuit,tanaka2005anomalous,tanaka2022theory} and spin-active interfaces \cite{huertas2002absolute,cottet2009spin,eschrig2009scattering, bergeret2012electronic,bergeret2012spin,eschrig2015general}.

 Boundary conditions can also  be described within the action formalism. For spin-independent boundary conditions actions historically have been derived both by constructing effective actions whose minimization yields known boundary conditions \cite{belzig2001full,nazarov2009quantum}, and based on microscopic foundation within the framework of the nonlinear sigma model (NLSM) \cite{feigelman2002weak,kamenev2023field}, leading to a quasiclassical action whose saddle point reproduces the kinetic transport equations and their boundary conditions. Actions for some special cases with spin-dependent boundary conditions have been derived phenomenologically \cite{mazanik2025interfacial,kokkeler2024nonreciprocal}.

In this article, we derive general  boundary conditions for spin-activeinterfaces with arbitrary transparency based on an effective boundary action, which we derive from a microscopic model.  The expressions are compact and the method is adaptable to a wide variety of interfaces.
In particular, we derive boundary conditions for hybrid interfaces featuring a generic interfacial exchange field and spin-dependent transmission eigenvalues. We show that the resulting expressions apply to both ferromagnetic and antiferromagnetic interfaces between different materials. Furthermore, we demonstrate how this approach can be extended to junctions where one material is in the clean limit and the other is diffusive. This allows us to describe the proximity effect of unconventional superconductors on dirty metals through a spin-activeinterface.
%Our method not only reproduces known boundary conditions in a compact and transparent way but also provides a systematic framework to combine multiple interfacial effects within a single boundary condition.

%We discuss three distinct things: (i) derivation of Tanaka-Nazarov via NLSM similar to Kamenev for Nazarov, (ii) generalization of those conditions interfaces with an intrinsic exchange field and spin-dependent tunneling at the same time (only one of them is rather trivial), (iii) boundary conditions via symmetry arguments.

The structure of this article is as follows. First, in Sec. \ref{Sec:main}, we summarize the main findings of our manuscript and present the new boundary conditions and the corresponding actions. 
In Sec. \ref{sec:TanakaNazarov} we show how the derivation presented in Ref.  \cite{kamenev2023field} can be generalized to describe interfaces between a dirty and a clean material, and thereby derive a generalized form of the Tanaka-Nazarov boundary conditions \cite{tanaka2003circuit,tanaka2005anomalous,tanaka2022theory}. Moreover, through the method of derivation, the results are also immediately applicable to interfaces with spin-dependent transmission. In Sec. \ref{Sec:spin-dependent} we generalize the Nazarov \cite{nazarov1999novel} and Tanaka-Nazarov boundary conditions \cite{tanaka2003circuit,tanaka2005anomalous,tanaka2022theory} to systems that contain an interfacial exchange field, and use the low transparency limit to relate them to those obtained in \cite{cottet2009spin}. Then, in Sec. \ref{sec:FCS} we show charge current conservation and derive the full counting statistics actions for charge, thermal, spin and spin-energy currents. In Sec. \ref{sec:anomalousproximity} we use the boundary conditions to describe the effect of a magnetic boundary on the anomalous proximity effect due to surface Andreev bound states.
In Sec. \ref{sec:JJ} we extend this analysis to contrast two different types of Josephson junctions, a junction between two spin-split superconductors and a junction between a spin-split superconductor and a triplet superconductor. We show that while both of these junctions have the potential to host anomalous phases and diode effects \cite{ando2020observation,nadeem2023superconducting,ilic2022theory,he2022phenomenological}, the relative orientations of the fields required to obtain these effects are considerably different. We explain this difference with the help of time-reversal symmetry.

%In Sec. (i) there is now a theory, but there is a point which should clearly be written better. Sec. (ii) is a derivation that is already in a relatively neat form.
%Sec. (iii) are only a few considerations, not of any use yet.

%In Kamenev's book, a derivation for the action of Nazarov's boundary conditions is presented. Here we show that the same formalism may be applied in case of one dirty and one clean material, and in this way the Tanaka-Nazarov boundary condition can be obtained. The same formalism may also be used for two clean materials, however, in this case no boundary condition can be obtained.

\section{Main results\label{Sec:main}}

\begin{figure*}
    \centering
    \includegraphics[width=17.2cm]{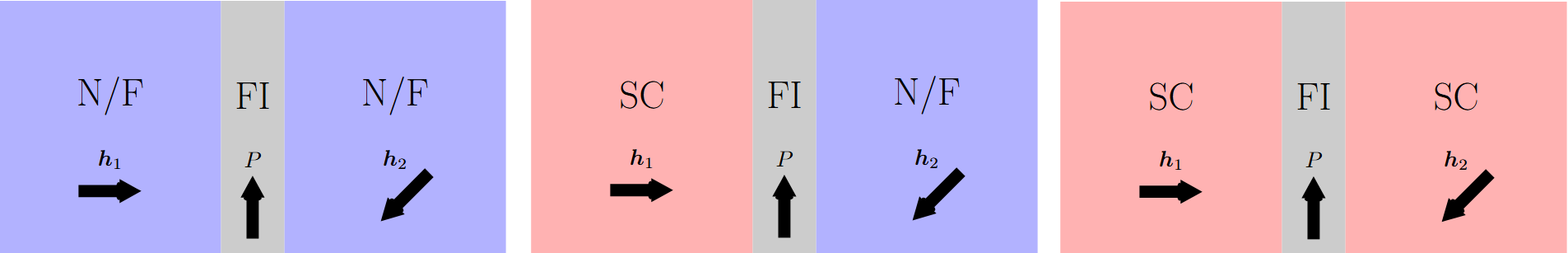}
    \caption{The action formalism can be used to derive boundary conditions for both materials for junctions between dirty normal metals, ferromagnets and (spin-split) superconductors with a spin-active barrier such as a ferromagnetic insulator. }
    \label{fig:Dirtyjunctions}
\end{figure*}

\begin{figure*}
    \centering
    \includegraphics[width=12.9cm]{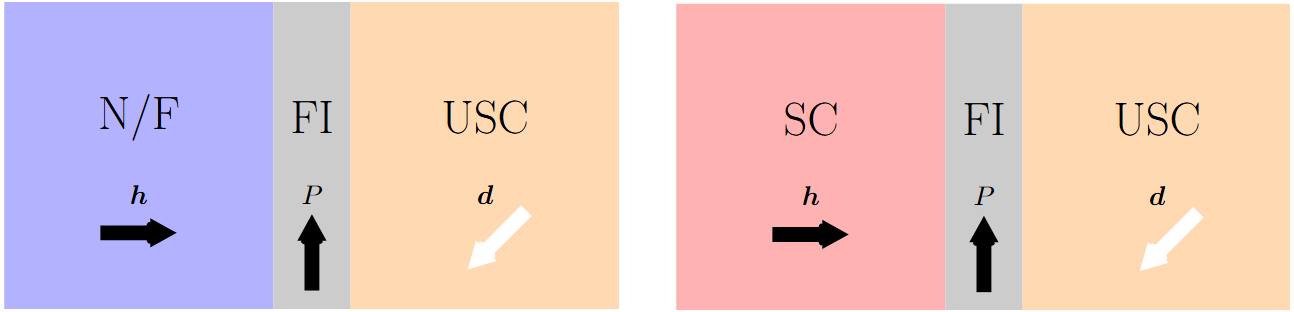}
    \caption{The action formalism can be used to derive boundary conditions for dirty materials for junctions consisting of a dirty material and a clean material, such as an unconventional (triplet) superconductor, connected via a spin-active barrier.}
    \label{fig:Cleanjunctions}
\end{figure*}
Before we delve into the details of the formalism we summarize the main findings of this manuscript. We derive the action for two types of junctions with spin-active barriers within the nonlinear sigma model (NLSM) formalism: spin-active barriers between two dirty materials, as illustrated in Fig.~\ref{fig:Dirtyjunctions}, or junctions in which one of the materials is clean, for example an unconventional superconductor, illustrated in Fig.~\ref{fig:Cleanjunctions}.

%For a spin-active barrier between two dirty materials, such as normal metals, ferromagnets, altermagnets or conventional superconductors, we derive the following boundary action, which can be expressed through the Q-matrices $Q_{L(R)}$ of the materials on the left (right) side of the junction. 
%\begin{align}
%     &iS_d[Q_L,Q_R]  =\text{Tr ln}(\mathbf{1}+i\pi\nu Q_{L}\vec{h}\cdot\vec{\sigma}\tau_{3}) + \text{Tr ln}\Big(\mathbf{1}\nonumber\\&+i\pi\nu Q_{R}\vec{h}\cdot\vec{\sigma}\tau_{3}+(\pi\nu)^2 Q_{R}W^{\dagger}(\mathbf{1}+i\pi\nu Q_{L}\vec{h}\cdot\vec{\sigma}\tau_{3})^{-1}Q_{L}W\Big)
%\end{align}

%\SB{This section can be shortened quite a bit. I would write a single action, let us say on the rhs, Eq. 1. We may denote the $Q_R$ as $\tilde Q_R$ and write in the text that $\tilde .$ denotes either the matrix  $Q$ which coincides with the Usadel $g$ at the saddle point or the Eilenberger $g$ depending on mometum parallel to the junction. This will be shown in section XXX.  The action for the left hand side may not be necessary to be shown. We then write the currents using the $\tilde Q$. I would say these summarized the main results of the paper. What do you think?}\TK{Agree, have done so.}

In general terms the total action of the junction can be written as:
\begin{align}
    S=S_L+S_R+S_T\; , 
    \end{align}
where $S_{L,R}$ are bulk actions of the left (L) and right (R) materials, and $S_T$ is the boundary action. 
We assume that the right material is a diffusive metal, so its bulk action $S_{R}[ Q_{R} ]$ is a functional of the matrix field $Q_{R}$, which is introduced to decouple the four-fermion term arising after disorder averaging~\cite{kamenev2023field}. Restricting $Q_R$ to describe massless, i.e.\ electronic diffusive, modes imposes the nonlinear constraint $Q_R^{2} = 1$. The manifold defined by this constraint is called the soft mode manifold.

As was shown in Appendix B of Ref. \cite{kokkeler2025quantum}, variation  of the  action with respect to $Q_R$ leads to 
\begin{align}
   &\delta S=\text{Tr}\int_{\Omega_{R}} dV\Bigg( \Big(\frac{\delta S_R}{\delta Q_R}-\partial_{j}\frac{\delta S_R}{\delta \partial_{j}Q_R}\Big)\delta Q_R\Bigg) \nonumber\\&+\text{Tr}\int_{\partial\Omega} dS\Bigg(\Big( n_{j}\frac{\delta S_R}{\delta \partial_{j}Q_R}+\frac{\delta S_T}{\delta Q_R}\Big)\delta Q_R\Bigg)\;,\label{eq:Action0}
\end{align}
where $\int_{\Omega_{R}} dV$ denotes integration over the right material, $\int_{\partial\Omega} dS$ denotes integration over the system boundary and $n_{j}$  the local normal vector to the boundary. 

The kinetic equation and its boundary condition are obtained via the saddle point condition, that is, by requiring that $\delta S = 0$ for any perturbation that respects the normalization condition $Q^2 = 1$ \cite{kamenev2023field}. In the bulk of the $R$ electrode, this follows from the first line of Eq.~(\ref{eq:Action0}) and leads to the Usadel equation, which, depending on the type of fields present (exchange, spin-orbit coupling, etc.), takes different forms~\cite{usadel1970generalized,bergeret2014spinorbit,bergeret2016manifestation,tokatly2017usadel,virtanen2021magnetoelectric,virtanen2022nonlinear,kokkeler2025universal,kokkeler2025quantum}. On the other hand, the boundary conditions for this equation are obtained by minimizing the action at the boundary, $\delta S|_{\delta\Omega}=0$, {\it i.e.} the second line of Eq.~(\ref{eq:Action0}) vanishes for any the perturbation in $Q$ that can be written  as $\delta Q = [\alpha,Q]$, where $\alpha$ is an arbitrary test function (see Appendix B of Ref. \cite{kokkeler2025quantum} for details). At the boundary one obtains, from the second line of Eq.~(\ref{eq:Action0}):
\begin{align}
    %\vec{n}_L\cdot\vec{J}_{L} = \frac{\partial iS_d}{\partial Q_L}\;,\\
    \vec{n}_R\cdot\vec{J}_R = i[Q,\frac{\delta S_T}{\delta Q_R}]\;,
    \label{eq:general_BC}
\end{align}
where the matrix current is defined as
\begin{align}
\label{eq:general_current}
J_{Rj} = i[Q,\frac{\delta S_R}{\delta \partial_{j}Q_R}]\;.
\end{align}
At the saddle point, the Q-matrix corresponds to the quasiclassical Green's function of the system. We denote the oppositely oriented outward normal vectors of the two materials by $\vec{n}_{L,R}$. 

Equation (\ref{eq:general_BC}) is a general expression that allows one to derive boundary conditions for the Usadel equation at any interface, provided the boundary action is specified. For the latter, we generalize the action introduced in Ref.~\cite{kamenev2023field} to include spin-dependent transmission coefficientes and an interfacial exchange field:
\begin{align}
     &iS_T[\tilde{Q}_L,Q_R]  = \sum_n\int_{\partial\Omega} dS\; \text{Tr ln}\Big(\mathbf{1}+i\pi\nu Q_{R}\vec{h}\cdot\vec{\sigma}\tau_{3}\nonumber\\&+(\pi\nu)^2 Q_{R}W_n^{\dagger}(\mathbf{1}+i\pi\nu \tilde{Q}_{L}\vec{h}\cdot\vec{\sigma}\tau_{3})^{-1}\tilde{Q}_{L}W_n\Big)\; . \label{eq:Action1}
\end{align}
Here, $\nu$ is the density of states of the right material, $\vec{h}$ the effective interfacial exchange field and $W_nW_n^{\dagger}$ is a matrix containing the transmission rates of the $n$th mode of the constriction. The transmission rates are spin-dependent. In general, $W_n$ can be parameterized as
\begin{align}
    W_n = t_n + u_n \vec{P}_n\cdot\vec{\sigma}\;,
\end{align}
where $\vec{P}_n$ is the spin-quantization axis of the $n$th mode, while the transmission rates of the mode are $|t_n+u_n|^2$ for spins parallel to $\vec{P}$ and $|t_n-u_n|^2$ for spins antiparallel to $\vec{P}$.

We use the notation $\tilde{Q}_{L}$ to emphasize that the left electrode is not necessarily diffusive. If the left material is diffusive, $\tilde{Q}_L$ is the Q-matrix of the left, in analogy to $Q_R$. As discussed in the next section, for clean materials the saddle point $\tilde{Q}_{L}$ corresponds to the quasiclassical Eilenberger Green function of the isolated left side, $g_{L}$. %In Eq.~(\ref{eq:Action1}), $\vec{h}$ denotes an 
The interfacial exchange field, $\vec{h}$, is finite at spin-active interfaces (see Section~\ref{Sec:spin-dependent}), and is here assumed to be homogeneous. 
%because we assume that the contact between the materials is small, and
%In Eq.~(\ref{eq:Action1}),
%$W$ is the tunneling matrix describing the interface between the two materials. 
If the boundary is with vacuum, then both $\vec{h}$ and $W$ vanish, and $S_{T} = 0$.
%\SB{ why do we include in $S_T$ the interfacial exchange only but not other terms as relaxation etc? We should comment on that, and cite our  papers and the paper  with Andrei.} \TK{Here we start from microscopics. Because there is a logarithm there is actually some spin-relaxation, but only one of the two types and with a specific relation to the exchange field term. This is because we assume the magnetic field is constant within the point contact, if you let go of this assumption things become ugly. However, because we have the logarithm, if we compare to the SMR-type BC we used before, $G_i$ (of course) and one type of relaxation, $G_r$, are there, only $G_s$ is missing, because that one requires some randomization of the interfacial spins. Next to this, the terms in the paper with Andrei are not there because we do not consider SOC here} It is important to emphasize that the matrix $W$ may have a non-trivial structure in spin space, hence the action (\ref{eq:Action1}) is a generalization of the one in Refs. \cite{nazarov1999novel,kamenev2023field}.

As detailed in Sec. \ref{Sec:spin-dependent} and the supplemental material, substitution of Eq.~(\ref{eq:Action1}) into Eq.~(\ref{eq:general_BC}) gives
\begin{align}
\vec{n}_R\cdot\vec{J}_R &=\sum_n[(1+X_nQ_{R})^{-1}X_n,Q_{R}]\;, \label{eq:CurrExpr1}
\end{align}
where 
\begin{align}
    X_n = i\pi\nu\vec{h}\cdot\sigma\tau_3+(\pi\nu)^{2}W_{n}^{\dagger}(\mathbf{1}+i\pi\nu\tilde{Q}_{L}\vec{h}\cdot\vec{\sigma}\tau_{3})^{-1}\tilde{Q}_{L}W_{n}\;.\label{eq:Currexpr4}
      %\vec{n}_L\cdot\vec{J}_{L}&=[(\mathbf{1}+\Tilde{X}Q_{L})^{-1}\Tilde{X},Q_{L}]\;,%\\
%\Tilde{X} &= i\vec{h}\cdot\sigma\tau_{3}+(\pi\nu)^{2}W^{\dagger}(\mathbf{1}+iQ_{R}\vec{h}\cdot\vec{\sigma}\tau_{3})^{-1}Q_{R}W\label{eq:Currexpr4}\;.
\end{align}

Expressions~(\ref{eq:CurrExpr1}-\ref{eq:Currexpr4}) together with Eq. ~(\ref{eq:general_current}) constitute the sought boundary condition and represent the main general result of this work. They describe junctions for any dirty materials with a magnetized constriction barrier between them within the quasiclassical approximation, as well as the the boundary between any dirty and clean material with a magnetized constriction within the quasiclassical approximation. Moreover, the action, Eq.~(\ref{eq:Action1}) can be used beyond the saddle point to calculate higher-order current correlations.

Notice that if  the left material is also diffusive, the boundary conditions for this material can be found by exchanging the left and right indices in the  expressions (\ref{eq:general_current}-\ref{eq:Currexpr4})  and taking the Hermitian conjugate of $W_{n}$. In the next sections, we derive this boundary condition and use it to illustrate different boundaries and physical situations.

\section{Derivation of Tanaka-Nazarov boundary conditions via NLSM}\label{sec:TanakaNazarov}

First,  we focus on the boundary between a clean and a diffusive metal. For illustrative purposes, we neglect the effective exchange field within the barrier within this section, that is, we keep $\vec{h = 0}$, but allow for spin-dependent transmission matrices $W$.
The influence of the exchange field is discussed in Sec. \ref{Sec:spin-dependent}.

We follow a similar derivation to the one presented in Ref. \cite{kamenev2023field} for all diffusive systems. 
%In the context of the nonlinear sigma model, one derives the  quasiclassical action for the   $Q$-matrix, 
%\cite{kamenev2023field}, 
 %which obeys the nonlinear constraint $Q^{2} = \mathbf{1}$. Additionally, when using Nambu space, additionally the charge conjugation constraint $Q_{t,t'} = \tau_{1}\sigma_{y}Q_{t',t}\sigma_{y}\tau_{1}$ is required. To find the kinetic equations and their boundary conditions, the action should be minimized on the soft-mode manifold defined by these two constraints.
%At the saddle point of the action, $Q$ can be identified with the quasiclassical Green's function $g$. The saddle point equations thus correspond to the dirty limit kinetic transport equations, the Usadel equation.
%This Usadel equation has to be equipped with boundary conditions decsribing hybrid interfaces. 
%which within the NLSM framework can be done with the help of a boundary action. For a tunneling barrier of arbitrary transparency between two dirty materials, this action has been derived in \cite{kamenev2023field}, and the resulting boundary conditions correspond to those of the Nazarov circuit theory. Here we follow a similar approach, but we focus on two different generalizations. 
%
%In this section we discuss  the boundary action in the case one of the materials is clean. 
%and use this to find a generalized version of the Tanaka-Nazarov boundary conditions. Next we explain how the boundary conditions change if the boundary becomes spin-dependent, that is, the transparency eigenvalues are different for opposite spins and there is a local exchange field at the boundary. 
Specifically, we consider a junction between a clean material  on the left (L), and a dirty material on the right (R).  We disregard electron-electron interaction in the electrodes such that the action of the system can be written as %\SB{are the imaginary $i$'s correct here? We shoudl mention time strcture, Keldysh etc.}:
%\SB{We need to say something about filed $Q$ etc. At least give some references and explain with words. Where can one find the strcture of $S_t$? What is $W$?}
\begin{align}
    e^{S} &= \int \mathcal{D}\psi_{L,R} \mathcal{D}\bar{\psi}_{L,R}\;e^{S_{L}+S_{T}+S_{R}}\;,\\
    S_{L}& = \int dV \int dt\; \bar{\psi}_{L}G_{L}^{-1}\psi_{L}\;,\\
    S_{R}& = \int dV \int dt\; \bar{\psi}_{R}(G_{R}^{-1}+\frac{i}{2\tau}Q)\psi_{R}\;,\\
    S_{T}& = \int dS\; \bar{\psi}_{L}W\psi_{R}+\text{h.c.}\;.
\end{align}
We denote the fermionic  bispionors as $\psi_{L,R}$, the microscopic Green's functions by $G_{L,R}$. The notation $\mathcal{D}$ is used to indicate path integrals over Grassmann variables, while the notation $d$ is used for integration over real variables.  The action for the right electrode contains  the disorder term $\frac{i}{2\tau}Q$, where $\tau^{-1}$ is the momentum scattering rate, and $Q$ is 
the matrix used to decouple the quartic Fermion terms due to disorder \cite{kamenev2023field}. The $Q$-matrix obeys the nonlinear constraint $Q^{2} = \mathbf{1}$. The  Nambu structure additionally imposes the charge conjugation constraint $Q_{t,t'} = \tau_{1}\sigma_{y}Q_{t',t}\sigma_{y}\tau_{1}$ \cite{kokkeler2025universal,kokkeler2025quantum}. At the saddle point of the action, $Q$ coincides  with the quasiclassical Green's function $g$. The saddle point equations thus correspond to the dirty limit kinetic transport equations, the Usadel equation \cite{usadel1970generalized}.
At the boundary, the spinors are connected with the help of the transmission matrix $W$, which depends on the in-plane momentum, contains a spin degree of freedom but is diagonal in Nambu space. The transmission matrix is restricted to the vicinity of the junction where electron wave functions overlap.

Because all terms in the actions are second order in the Fermion fields $\psi_{L,R}$, we may use the rules for Gaussian integration for Grassmann variables to integrate them out and obtain
\begin{align}
    iS = -\text{Tr}Q^{2}+\text{Tr ln}\begin{bmatrix}
        G_{L}^{-1}&W\\W^{\dagger}&G_{R}^{-1}+\frac{i}{2\tau}Q
    \end{bmatrix}\;.\label{eq:ActionToStartFrom}
\end{align}
Here, and in what follows for simplicity of notation we absorb integration over space and time coordinates into Tr, while traces over Nambu-spin-Keldysh space are indicated by tr.
Following Ref.\cite{kamenev2023field}, we can write $\mathbf{G}_{R}^{-1} = G_{R}^{-1}+\frac{i}{2\tau}Q$ and $\mathbf{G}_{L}^{-1} = G_{L}^{-1}$ and express the second part of the action as $iS = -\text{Tr}Q^2+iS'$, with
%\SB{we shoudl distinguish $S$'s. Should we denote it  $S_T$ as Kamenev?} 
\begin{align}\label{eq:Plus}
    iS' = \text{Tr ln}\begin{bmatrix}
        \mathbf{G}_{L}^{-1}&0\\0&\mathbf{G}_{R}^{-1}
    \end{bmatrix}+\text{Tr ln}\left(1+\begin{bmatrix}
        0&\mathbf{G}_{L}W\\\mathbf{G}_{R}W^{\dagger}&0
    \end{bmatrix}\right)\;,
\end{align}
%which has  the same structure as the expression in Ref. \cite{kamenev2023field} for two dirty superconductors, the only difference being the definition of $\mathbf{G}_{L}^{-1}$.
Using the equality $\text{Tr ln} A = \text{Tr ln} U^{-1}AU$ for any matrix $A$ and any invertible matrix we find that,  exploiting this relation with $U$ equal to  the third Pauli matrix in $L,R$ space,  the action of Eq.~(\ref{eq:Plus}) can be also written as
\begin{align}\label{eq:minus}
        iS' = \text{Tr ln}\begin{bmatrix}
        \mathbf{G}_{L}^{-1}&0\\0&\mathbf{G}_{R}^{-1}
    \end{bmatrix}+\text{Tr ln}\left(1-\begin{bmatrix}
        0&\mathbf{G}_{L}W\\\mathbf{G}_{R}W^{\dagger}&0
    \end{bmatrix}\right)\;,
\end{align}
Taking the half sum  of Eqs.~(\ref{eq:Plus}) and (\ref{eq:minus}), we obtain.
\begin{align}
    iS' &= \frac{1}{2}\Bigg(\text{Tr ln}\begin{bmatrix}
        \mathbf{G}_{L}^{-1}&0\\0&\mathbf{G}_{R}^{-1}
    \end{bmatrix}\nonumber\\&+\text{Tr ln}\left(1+\begin{bmatrix}
        0&\mathbf{G}_{L}W\\\mathbf{G}_{R}W^{\dagger}&0
    \end{bmatrix}\right)+\text{Tr ln}\begin{bmatrix}
        \mathbf{G}_{L}^{-1}&0\\0&\mathbf{G}_{R}^{-1}
    \end{bmatrix}\nonumber\\&+\text{Tr ln}\left(1-\begin{bmatrix}
        0&\mathbf{G}_{L}W\\\mathbf{G}_{R}W^{\dagger}&0
    \end{bmatrix}\right)\Bigg)\nonumber\\
    & = \text{Tr ln}\begin{bmatrix}
        \mathbf{G}_{L}^{-1}&0\\0&\mathbf{G}_{R}^{-1}
    \end{bmatrix}\nonumber\\&+\text{Tr ln}\left(1-\begin{bmatrix}
        \mathbf{G}_{L}W\mathbf{G}_{R}W^{\dagger}&0\\0&\mathbf{G}_{R}W^{\dagger}\mathbf{G}_{L}W
    \end{bmatrix}\right)\;,
\end{align}

%\SB{next discussion has to be polished and probably shorten. I would emphasize that $W$ are restricted to the interface. The original Nazarov BC is valid for interfaces with several (even infinte) conduction channels. }
The transmission matrices are only nonzero at the interface. For illustrative purposes, we consider a configuration similar to \cite{kamenev2023field} and assume that the two materials are connected by a constriction that is much smaller than the mesoscopic scales of the problem, so that we may approximate the spatial dependence by a delta function within the quasiclassical approximation, but larger than the atomic size so that multiple channels exist.

Unlike the Nazarov boundary condition discussed in \cite{kamenev2023field}, we can in general not replace $G_{L}$ by a momentum isotropic $Q$-matrix, since the left electrode is clean. %$Q$-matrix does not exist in the clean superconductor. 
 The boundary condition becomes dependent on whether the interface is clean or dirty. Here we  consider a perfectly clean interface, so that the in-plane momentum is conserved. To this end, we assume the diameter of the contact interface is large compared to the atomic scale, so that we can have multiple channels distinguished by the in-plane momentum of the modes, while it is at most of the order of the mean free path 
 on the dirty side, so that we may treat the interface as local in the nonlinear sigma model.
Thus, we can express the boundary condition in terms of the $Q$-matrix on the dirty side, while for the clean side we use the clean Green's function expressed as a function of the center of mass coordinate $R$, the $x$-component of the difference coordinate, $x-x'$, and the $y,z$ components of momentum $p_{y,z}$:
%that is, %we require $(G_{L}(R = 0,x-x' = 0^{+},p_{y},p_{z})+G_{L}(R = 0,x-x' = 0^{-},p_{y},p_{z}))/2$, where $R$ is the center of mass . In what follows we use the short-hand notation $G_{L}(r+r' = 0,x-x' = 0,p_{y},p_{z})$. It can be calculated as
\begin{align}
    &\mathbf{G}_{L}(R = 0,x-x' = 0,p_{y},p_{z}) \nonumber\\&= \frac{1}{\hbar}\int dp_{x} \mathbf{G}_{L}(R =0,p_{x},p_{y},p_{z})\approx -i \tilde{\nu}(p_y,p_z) g_{L}(0,p_{y},p_{z})\;,
\end{align}
%\TK{Check if it is the sum of $G$ evaluated at $x-x' = \pm \varepsilon$}
where $g_{L}$ is related by a density of states dependent prefactor to the angular averaged quasiclassical Green's function and $\tilde{\nu}$ is the density of states along a specific in-plane momentum. Since such prefactor can be absorbed in the transmission rate, we may take $g_{L}$ to be the actual quasiclassical Green's function.

%This quasiclassical Green's function  can be solved using the Eilenberger equation, the condition $r+r' = 0$ stems from the requirement of the pinhole being much smaller than the coherence length, $x-x' = 0$ from the fact that the interface normal is along the x-direction and $p_{y},p_{z}$ from the fact that the pin-hole is much larger than the atomic scale, so that the sum over all modes becomes an integration over all modes characterized by the conserved in-plane momentum.\SB{why is the conservation of momenum here important? Tunneling does not depend on momentum...}% It may be written as $g_{L}(0,\vec{n}) = g_{L}(0,-\vec{n})$, where the latter comes from the no-current condition for the clean material, that arises because the current in this formalism is at most of order $\tau$ and hence negligible in the superconductor.  %\TK{There is a twist if we consider multiple modes. If one naively takes the angular averaged one, then we should have the angular averaged Green's function in the denominator as well. However, Kamenev also separates the channels immediately, which he needs to because transmission is not the same for all modes (in his case it does not affect the outcome though). He assumes $W$ is diagonal in the channel basis to get it outside. This is strange because $Q$ has no channel index, while $W$ does. It seems that $G(r,r)$ also still must have channel indices, where $n$ is the channel index. However, then is much harder to transform to $Q$ (you have to assume all channels are the same)}

We now obtain an action for the matrix field $Q_R$:
\begin{align}\label{eq:ActionW}
    S'[Q_R] &= S_{R}[Q_R]+S_T[Q_R]\;,\\
    S_T[Q_R] &= \sum_n\text{Tr ln}\left(1+(\pi\nu)^2W_{n}^{\dagger}g_{L}W_{n}Q_R\right)|_{T}\;.
\end{align}
where the summation is over the different channels of the constriction, characterized by in-plane momentum $(p_x,p_y) = (p_{x,n},p_{y,n})$, and we defined the shorthand $g_{Ln} = g_{L}(p_{x,n},p_{y,n})$ and $W_{n} = W(p_{x,n},p_{y,n})$. $S_{R}[Q_R]$ is the action in the right material in the absence of transmission, $\nu$ is the momentum direction averaged density of states and the notation $|_{T}$ denotes it should only be evaluated at the tunnel boundary. This implies the following boundary condition at this boundary via variation of $Q_{R}$:
\begin{align}\label{eq:JW}
    \vec{n}\cdot\vec{J} = \sum_{n} [(1+(\pi\nu)^2W_{n}^{\dagger}g_{Ln}W_{n}Q_R)^{-1}(\pi\nu)^2W_{n}^{\dagger}g_{L}W_{n},Q_R]\;,
\end{align}
where $J$ is the current that follows from the bulk equation, see Eq.~(\ref{eq:general_current}).

 Thus, the boundary conditions for the case in which the left material is clean can be obtained from those in which both are in the dirty limit, Ref. \cite{kamenev2023field}, by replacing the Q-matrix on the left hand side by the momentum integrated surface Green's function $g_L$. In the case both of the materials are dirty and the interface is not spin-active, the Nazarov boundary conditions are recovered by defining $ T_{1n}\mathbf{1} = (\pi \nu W_{n})^2$,  and substitution of $g_L$ by $Q_L$. Since $T_1$ in the Nazarov boundary condition is related to the transparency $\tilde{T}$ via 
$\tilde{T} = \frac{4T_{1}}{(1+T_{1})^2}
$, the eigenvalues of $W_n$ are directly related to the spin-dependent transparency via the equivalent formula.

%\SB{in the last expression there are things to be clarify: 1) for the current we can refer to the previous section after we add the general expression. 2) the l.h.s of this expressions does not depend on $p_{y,z}$ however the r.h.s does! Should we take the average on the r.h.s?}

\subsection{Connection to Tanaka-Nazarov boundary conditions}
The boundary equation in Eq.~(\ref{eq:JW}) is valid for a wide variety of interfaces and left materials. Specifically, it does not require the left material to be in the quasiclassical regime, since $g_L$ is defined as an integral over the microscopic Green's function that can always be evaluated. 
In case the material on the left side is homogeneous in the quasiclassical regime and the interface is clean, our equations should reproduce the Tanaka-Nazarov boundary conditions \cite{tanaka2003circuit}. We now show that this is indeed the case. 

The resulting boundary condition contains the variable $-i\tilde{\nu} (p_y,p_z) g_{L}(0,p_{y},p_{z}) = \frac{1}{\hbar}\int dp_{x}\vec{G}_{L}(R = 0,p_{x},p_{y},p_{z})$, where $\tilde{\nu}$ is the density of states along a specific momentum direction. Thus, in the quasiclassical limit $g_L$ corresponds to a quasiclassical Green's function. Importantly, since $g_{L}$ is calculated from $\vec{G}_{L}$ only, it has no notion of the right-hand side, and is thus the surface Green's function of the material in contact with vacuum. This is a consequence of the initial approximation that is inherent to writing the action in two fermion terms. If the left hand side is well described using the quasiclassical approximation, we may find an explicit expression for $g_L$ using the quasiclassical Eilenberger equation on the clean side.

Now suppose that on the left material all material parameters are homogeneous. The Eilenberger equations for the incoming and outgoing modes may be written as
\begin{align}
    n_{Fx}\partial_{x}g_{L+} &= \alpha_{1}[\Lambda_{1},g_{L+}]\;,\label{eq:gLpbulk}\\
    n_{Fx}\partial_{x}g_{L-} &= -\alpha_{2}[\Lambda_{2},g_{L-}]\label{eq:gLmbulk}\;,
\end{align}
where $\Lambda_{1}^{2} = \Lambda_{2}^{2} = 1$ correspond to the bulk quasiclassical Green's functions for the incoming and outgoing momenta and $\alpha_{1,2}$ are real scalars that may depend on energy and in-plane momentum. For simplicity of notation, we have dropped the channel index notation of $g$.
If $L$ acts as a reservoir, at the boundary we may impose the no-current condition $g_{L+}(x = 0,p_{\parallel}) = g_{L-}(x = 0,p_{\parallel})=:C(p_{\parallel})$, where $C$ is the surface Green's function. %\SB{I would write dependece on $p_\parallel$ as well}
Because $g_{L+}$ can not diverge as $x\xrightarrow{}-\infty$, we must have that
\begin{align}
    (1-\Lambda_{1})(C+1) = 0\label{eq:Cc1}\;,
\end{align}
as can be verified by considering the basis in which $\Lambda_{1}$ is diagonal.
Moreover, since $g_{L-}$ can not blow up either, we have an extra condition,
\begin{align}
    (1+\Lambda_{2})(C-1) = 0\label{eq:Cc2}\;.
\end{align}
These two conditions define a unique solution, as detailed in Appendix \ref{sec:CalculatingC}.
Defining $H_{\pm} = \frac{1}{2}(\Lambda_{1}\pm\Lambda_{2})$, we find
\begin{align}\label{eq:Cexpression}
    C&= H_{+}^{-1}(1-H_{-})\;.
\end{align}

In terms of $C$, the boundary condition reads \begin{align}\label{eq:JWC}
    \vec{n}\cdot\vec{J} = \sum_{n} [(1+(\pi\nu)^2W_n^{\dagger}CW_nQ_R)^{-1}(\pi\nu)^2W_n^{\dagger}CW_n,Q_R]\;,
\end{align}
where $C$ is understood to be evaluated at $(p_{y},p_z)_n$.
In the limit of spin-independent interfaces (so that $\pi\nu W = \sqrt{T_{1n}}\mathbf{1}$), this simplifies to
\begin{align}
    \vec{n}\cdot\vec{J} &= \sum_{n} [(1+T_{1n}CQ_R)^{-1}T_{1n} C,Q_R] \nonumber\\& =\sum_{n} (1+T_{1n}CQ_R)^{-1}T_{1n} CQ_R-T_{1n}Q_R C (1+T_{1n} Q_R C)^{-1} \nonumber\\&= \sum_{n}(T_{1n}1+T_{1n}^2 + T_{1n}\{C,Q_R\})^{-1}CQ_R\nonumber\\&- T_{1n}Q_R C (1+T_{1n}^2 + T_{1n}\{C,Q_R\})^{-1}\nonumber\\&= \sum_{n}(1+T_{1n}^2 + T_{1n}\{C,Q_R\})^{-1}T_{1n}[C,Q_R]\;.
\end{align}

This is exactly the same expression as appears in the most recent formulation of Tanaka-Nazarov boundary conditions \cite{tanaka2022theory}. 

\section{Interfacial exchange field \label{Sec:spin-dependent}}
In this section, we discuss the boundary conditions including an interfacial exchange field. 
In principle, such boundary conditions have been derived for diffusive systems in previous works~\cite{cottet2009spin,bergeret2012spin,eschrig2015general}. 
Here, however, we present a compact and simpler  way of deriving them using our general considerations from Sec.~\ref{Sec:main}. 
In particular, all equations hold both in the case of two diffusive materials and in the case of a single diffusive material. 
In what follows, we focus only on the case of two diffusive materials, as in previous works~\cite{nazarov1999novel,huertas2002absolute,cottet2009spin,bergeret2012spin}. 
Nevertheless, it should be kept in mind,  as shown in the previous Section, that the boundary conditions in case the left hand side is clean can be obtained by replacing $Q_{L}$ with the corresponding quasiclassical Green’s function.

At magnetic interfaces, two processes need to be taken into account. 
On the one hand, the transmission matrix $W_{n}$ connecting the two materials can be spin-dependent, which leads, for example,  to spin-filtering effects~\cite{meservey1970magnetic,tedrow1971spin,julliere1975tunneling,moodera1988electron,hao1990spinfilter,moodera1993variation}. 
On the other hand, the magnetic moments at the interface may interact with the conduction electrons of the $L$ and $R$ materials, giving rise to spin torque and relaxation effects \cite{brataas2001spin,huertas2002theory,chen2013theory,huang2012transport,vlietstra2013spin,nakayama2013spin,dejene2015control,zhang2019theory}. %~\cite{smr,brataas,nazarov,bauer,...}. 
These processes may occur even in the absence of transmission, for instance at the interface between a metal and a ferromagnetic interface (FI), which can be realized as a layer of a ferromagnetic insulator or a thin ferromagnetic metal, which is described by the interfacial exchange field ${\bf h}$ in Eq.~(\ref{eq:Action0}).% \SB{as mentioned before we may include also the relaxation}

In the absence of an interfacial exchange field, i.e. $
\vec{h}=0$,  Eq.~(\ref{eq:JW}) holds. 
%\SB{one should define $T_{1n}$ and briefly explain how to obtain this expression} \TK{It was a typo, it is the factor that is there if you do not write the $W$'s}
In the limit $T_{1n}\ll 1$,  this reduces to the expression presented in  Ref. \cite{bergeret2012spin}, and widely used to describe magnetic tunneling barriers \cite{heikkila2019thermal}.
In the case of a thick FI,  transmission can be neglected.  In that case the two materials may be treated separately, and we find
that a term $[\delta(r)\vec{h}\cdot\sigma\tau_{3},Q_{L,R}]$ appears in the Usadel equations on the left and right side, which can be incorporated in the boundary condition.  The derivation of the boundary conditions for materials in which the two effects coexist requires some careful analysis.  We start from the action that is obtained by adding the interfacial exchange field to Eq. (\ref{eq:ActionToStartFrom}), %\SB{can we refer to some previosu expression? I would focus only on the interfae action, what we call $S_T$ in section II, which would coincide with equation A1.}
\begin{align}
iS& =
\text{Tr ln}\begin{bmatrix}
        \mathbf{G}_{L}^{-1}&0\\0&\mathbf{G}_{R}^{-1}
    \end{bmatrix}\nonumber\\&+\text{Tr ln}\left(1+\delta(r)\begin{bmatrix}
        \mathbf{G}_{L}\vec{h}\cdot\vec{\sigma}\tau_{3}&\mathbf{G}_{L}W_{n}\\\mathbf{G}_{R}W_{n}^{\dagger}&\mathbf{G}_{R}\vec{h}\cdot\vec{\sigma}\tau_{3}
    \end{bmatrix}\right)\;.\label{eq:ActionGeneralBC}
\end{align}    
In contrast to Eq. (\ref{eq:ActionToStartFrom}),the second term now contains both diagonal and off-diagonal terms and hence the procedure laid out in the previous section can not be followed through to the end. 
\begin{comment}
\subsubsection{Special cases}
In the special case in which $\vec{h} = 0$, the derivation can be followed very similarly to the one in Kamenev. Indeed, doing the expansion we note that the trace of the first order term is zero, while the second term is diagonal in $L,R$ space and reads $\begin{bmatrix}
    G_{L}W_{n}G_{R}W_{n}^{\dagger}&0\\0&G_{R}W_{n}^{\dagger}G_{L}W_{n}
\end{bmatrix}$.
Since we are taking the trace, the difference in order of $G_{L}W_{n}$ and $G_{R}W_{n}^{\dagger}$ is irrelevant and we arrive at
\begin{align}
    iS = iS_{0}+ \text{Tr ln}(1-G_{L}W_{n}G_{R}W_{n}^{\dagger}),
\end{align}
which as required is of a similar structure as Nazarov's equation.  W_{n}e may again substitiute $G_{L,R}$ by $i\pi\nu Q_{L,R}$ and arrive at
\begin{align}
    \text{Tr ln }(1+\pi^{2}\nu^{2}Q_{L}W_{n}Q_{R}W_{n}^{\dagger}).
\end{align}
The only change with the conventional case is that $W_{n}$ is now allowed to have matrix structure. In the case of spin-dependent scatterers we may decouple the channels again and arrive at the current
\begin{align}
    \sum_{n}\text{Tr}(1+Q_{L}W_{n}Q_{R}W^{\dagger})^{-1}[Q_{L},W_{n}Q_{R}W_{n}^{\dagger}].
\end{align}
In the limit of low transparencies this reduces to the heuristically derived expressions of Sebastian a long time ago.
Another special case is where $W = 0$. In this case the materials are completely decoupled and the boundary conditions obtain a term $[\vec{h\cdot\sigma},Q_{L,R}]$, further nothing really happens.
\end{comment}
%\subsection{General expression}
As shown in Appendix \ref{sec:GeneralExpression}, the most general expression for the current on the right side of the interface,  derived from the action in Eq.~(\ref{eq:ActionGeneralBC}) reads %\SB{The current or the r.h.s of the B.C. It is convenient to keep the notation of Eq.(\ref{eq:general_BC}). In other words susbtitution of $I_j$ by $\vec{n_j}\vec{J_j}$. Is it correct?}
\begin{align}
    \vec{n}&\cdot \vec{J}_R =\sum_n [(1+X_nQ_{R})^{-1}X_n,Q_{R}] \label{eq:magn_IR}\\
    X_n& = i\pi\nu\vec{h}\cdot\sigma\tau_3+(\pi\nu)^{2}W_{n}^{\dagger}(\mathbf{1}+i\pi\nu Q_{L}\vec{h}\cdot\vec{\sigma}\tau_{3})^{-1}Q_{L}W_{n}
    \label{eq:magn_X}\;,
\end{align}
where we used the notation $\vec{J}_R$ to emphasize it is the matrix current on the left side of the boundary.
 It is insightful to note this expression can also be written as
\begin{align}
    \vec{n}&\cdot \vec{J}_R = \sum_n(1+X_nQ_{R})^{-1}X_nQ_{R}-(1+Q_{R}X_n)^{-1}Q_{R}X_n\nonumber\\
    & = (1+\{X_n,Q_{R}\}+Q_{R}X_n^{2}Q_{R})^{-1}(X_nQ_{R}+Q_{R}X_n^{2}Q_{R})\nonumber\\&-(1+\{X_n,Q_{R}\}+X_n^{2})^{-1}(Q_{R}X_n+X_n^{2})\;.
\end{align}
In the absence of  magnetic interaction, $\vec{h} = 0$ and $W_{n}$ has no spin dependence, i.e. $(\pi\nu W_{n})^2 = T_{1n}\mathbf{1}$. Therefore it automatically commutes with $Q_{L,R}$. Thus, $X = T_{1n}Q_{L}$ and $X_n^{2} = T_{1n}^{2}\mathbf{1} = Q_{R}X_n^{2}Q_{R}$ is a scalar. In that case, we obtain the well known Nazarov boundary condition \cite{nazarov1999novel}:
\begin{align}
    \vec{n}\cdot \vec{J}_R = \sum_n (1+T_{1n}^{2}+T_{1n}\{Q_{L},Q_{R}\})^{-1}T_{1n}[Q_{L},Q_{R}]\;.
\end{align}
Note that the condition that allows for such a simple form is that the second order in $X$ terms squares to unity, something that we can not guarantee for general spin-active interfaces. %in the general case \SB{what is the general case?}.
For the left material  
%use the other expression for current. Next we can use a completely similar technique as for $Q_{R}$, since that expression contains only a single $Q_{L}$. Realizing that the 
one can obtain a similar expression
by switching the $L,R$ indices and interchanging $W_{n}$ and $W_{n}^{\dagger}$. In particular, Eqs.~(\ref{eq:magn_IR}--\ref{eq:magn_X}) transform to
\begin{align}\label{eq:JL}
    \vec{n}\cdot \vec{J}_L&=[(\mathbf{1}+\Tilde{X}_nQ_{L})^{-1}\Tilde{X}_n,Q_{L}]\;,
\end{align}
with 
\begin{align}
    \Tilde{X}_n = i\pi\nu\vec{h}\cdot\sigma\tau_{3}+(\pi\nu)^{2}W_{n}(\mathbf{1}+i\pi\nu Q_{R}\vec{h}\cdot\vec{\sigma}\tau_{3})^{-1}Q_{R}W_{n}^{\dagger}\;.
\end{align}

We note that, in general,  the matrix currents on the two sides of the interface Eqs.~(\ref{eq:magn_IR},~\ref{eq:JL}), are not necessarily the same, because in spin-active interfaces the spin current is in general not conserved.  This reflects that the boundary exerts a torque on the spin through the exchange field and the spin-dependent transmission.  Charge current, in contrast,  is obviously conserved, as elaborated in Sec. \ref{sec:FCS}.
%\SB{even if $h$ is zero I think a non trivila $W$ will lead to a non-conservation, right?}
%Indeed, as an example consider the case $W = 0,\vec{h}\neq 0$ and two identical materials on each side. In that case the system is mirrorsymmetric, and there is a spin current flowing into both electrodes originating from this boundary, it acts as a source. Thus, $I_{L} = -I_{R}$ for this case instead of $I_{L} = I_{R}$ as in the conventional Nazarov's boundary condition. %The expressions above allow even for $|I_{L}|\neq|I_{R}|$, due to the contribution of both currents flowing from left to right or vice versa and of currents sourced by the magnetic field at the boundary. %\SB{magnetic field at the boundary?}

\subsection{Low transparency junctions}
%With this, we have derived tractable expressions for the matrix current through spin-dependent boundary conditions with arbitrary transparencies%\SBedit{
%, Eq.~(\ref{eq:magn_IR}). 
In several experiments interfaces are low-transmissive, and the boundary conditions can be simplified considerably, as we show in this section.
In the low-transparency limit we may keep in Eq.~(\ref{eq:magn_X}) only terms up to first order in transparency, i.e. one pair of $W_{n},W_{n}^{\dagger}$, and keeping of at most first order in $\vec{h}$.
We then obtain
\begin{align}
    X_n&\approx i\pi\nu\vec{h}\cdot\vec{\sigma}\tau_{3}+(\pi\nu)^{2}W_{n}^{\dagger}Q_{L}W_{n}\nonumber\\&-i(\pi\nu)^{3}W_{n}^{\dagger}Q_{L}\vec{h}\cdot\vec{\sigma}\tau_{3}Q_{L}W_{n}\;,
    \end{align}
    and hence for the  matrix current at R, $I = \sum_n I_n$, where
    \begin{align}
    I_n &= [(1+X_nQ_{R})^{-1}X_n,Q_{R}]\approx [X_n,Q_{R}]-[X_nQ_{R}X_n,Q_{R}]\nonumber\\& \approx i\pi\nu[\vec{h}\cdot\vec{\sigma}\tau_{3},Q_{R}]+(\pi\nu)^{2}[W_{n}^{\dagger}Q_{L}W_{n},Q_{R}]\nonumber\\&-i(\pi\nu)^{3}[W_{n}^{\dagger}Q_{L}\vec{h}\cdot\vec{\sigma}\tau_{3}Q_{L}W_{n},Q_{R}]\nonumber\\&-i(\pi\nu)^{3}[\vec{h}\cdot\vec{\sigma}\tau_{3}Q_{R}W_{n}^{\dagger}Q_{L}W_{n}+W_{n}^{\dagger}Q_{L}W_{n}Q_{R}\vec{h}\cdot\vec{\sigma}\tau_{3},Q_{R}]\;.
\end{align}
Writing \begin{align}W_{n} = t_n + u_n \tau_{3}\,\vec{P}\cdot\vec{\sigma}\;,\label{eq:Wspinfilter}\end{align} where $\vec{P}$ is a unit vector parallel to the barrier polarization, and neglecting terms of order $O(u_n^{2})$, the last expression can be written as
\begin{align}
    I&\approx i\pi\nu[\vec{h}\cdot\vec{\sigma}\tau_{3},Q_{R}]+(\pi\nu t_n)^{2}[Q_{L},Q_{R}]\nonumber\\&+(\pi\nu)^{2}t_nu_n[\{Q_{L},\vec{P}\cdot\vec{\sigma}\}]-i(\pi\nu)^{3}t_n^2[Q_{L}\vec{h}\cdot\vec{\sigma}\tau_{3}Q_{L},Q_{R}]\nonumber\\&-i(\pi\nu)^{3}t_n^2[\vec{h}\cdot\vec{\sigma}\tau_{3}Q_{R}Q_{L}+Q_{L}Q_{R}\vec{h}\cdot\vec{\sigma}\tau_{3},Q_{R}]\;.
\end{align}
This expression coincides with the boundary condition derived in Refs. \cite{cottet2009spin,eschrig2015general} to lowest orders in transparency, by making following identifications: the first term and part of the fourth term can be regrouped into $-iG_\phi [\hat{\vec{h}}\cdot\vec{\sigma}\tau_3, Q_R]$, where  $G_{\phi} = -\sum_n\pi\nu(1-(\pi\nu t_n)^{2})|\vec{h}|$, which leads to spin-triplet mixing and is present even if the layer is so thick there is no transmission, i.e. for a ferromagnetic insulator.  In this case it represents the $G_i$ term in SMR boundary conditions \cite{brataas2001spin,zhang2019theory,kokkeler2024nonreciprocal}. The remainder of the fourth term reads $-iG_{\chi}^{L}[Q_L[\hat{\vec{h}}\cdot\vec{\sigma}\tau_3,Q_L],Q_R]$ by identifying $G_{\chi}^L =\sum_n (\pi\nu)^3t_n^2 |\vec{h}|$ and corresponds a modification of the singlet-triplet mixing by the material on the other side of the barrier.
The second term corresponds to the usual tunnelling term by identifying $G_{T} = \sum_n(\pi\nu t_n)^{2}$. This is the only term remaining for spin-independent boundaries, and in that case corresponds to the Kuprianov-Luckichev boundary condition \cite{kuprianov1988influence}.  The third term corresponds to the magneto-resistance term via $G_{MR} = \sum_n (\pi\nu)^{2}t_nu_n$, while the fifth and last term appears because on a microscopic scale the correlations of the right material penetrate on the left and affect the exchange field there, with $G^{R}_{\chi} = \sum_n(\pi\nu )^{3}t_n^{2}|\vec{h}|$. Thus,   the spin-dependent phase is directly correlated to the strength of the interfacial exchange field.   If $\vec{h}\xrightarrow{}0$, our expressions recover those for a low spin-dependent tunnel barrier, derived heuristically in Ref. \cite{bergeret2012spin}, and used in several works, see e.g. \cite{silaev2015spin,heikkila2019thermal}.

To higher order in $W_{n},W_{n}^{\dagger}$ one obtains transparency corrections, while to second order in $\vec{h}$ one obtains relaxation  terms, specifically containing the  spin-mixing conductance at the boundary between a metal and  magnetic insulator \cite{brataas2001spin,zhang2019theory,kokkeler2024nonreciprocal}. %We note that the spin-sink conductance \cite{dejene2015control} does not appear in this particular model because of the absence of randomization of spin at the interface.
%\SB{the relaxation, I guess is the $G_r$ right? Maybe it is worth to say this. What about Eschrig et al? Do they have all orders in $W$? }\TK{Agree regarding first point, also made connection to $G_i$ in the previous paragraph. Regarding the second question: In both \cite{cottet2009spin,eschrig2015general} they only consider the case in which both sides are dirty. In \cite{eschrig2015general} they consider only specific cases such as low-transparency, FI or no spin-dependence at all. The discussion from Eq. 36 to Eq. 47 is for the fully generic case, but it expresses everything in effective $t$-matrices that need to be solved through some equation and Green's functions $g^{i,o}_{1,2}$ that still have incoming and outgoing degrees of freedom which means that it is not directly comparable, to compare we would need to rewrite their equation to something more compact, but since they do not do this it is probably not so easy. In \cite{cottet2009spin} they have a general expression, but it contains a matrix $\Sigma$ which is 1 or -1 depending on whether a mode has positive or negative $k_x$ (again the incoming/outgoing degree of freedom), so we can not really compare to that. Next to this, they have an expression without this $\Sigma$-matrix in which they keep all orders in what they call $T_{1n}$ (the transparency averaged over spin-directions), but already do an expansion in what they call $\delta X$, a non-trivial quantity that as far as I understand corresponds to an expansion in the different transparencies for different spins. }
\section{ Full counting statistics}\label{sec:FCS}

In this section, 
we derive the full counting statistics action, the action that can be used to compute all moments of the charge current \cite{levitov1996electron,belzig2001fullSN}. It was previously derived for the Nazarov boundary condition \cite{belzig2001full} and for the Tanaka-Nazarov boundary condition \cite{kokkeler2025full} by constructing the full counting statistics action consistent with the boundary condition. Here we show how to derive it  directly from the NLSM, including the spin dependence.
Along with this derivation, we show that our boundary conditions satisfy charge current conservation.

%An obvious requirement for the boundary condition is that charge currents are conserved at the boundary. In this section we show that our boundary condition satisfies this requirement. Along with it, we also obtain the full counting statistics action, the action that can be used to compute all moments of the charge current \cite{levitov1996electron,belzig2001fullSN}. It was previously derived for the Nazarov boundary condition \cite{belzig2001full} and for the Tanaka-Nazarov boundary condition \cite{kokkeler2025full} by constructing the full counting statistics action consistent with the boundary condition. Here we show how to derive the full counting statistics directly from the NLSM, including the spin dependence.

In general, we know that the matrix currents are not conserved. Indeed, denoting $\vec{n} = \vec{n}_L = -\vec{n}_R$, we have
\begin{align}
    \vec{n}\cdot \vec{\mathcal{J}}_L = i[Q,\frac{\delta S_T}{\delta Q_L}]\;,\\
    \vec{n}\cdot \vec{\mathcal{J}}_R = -i[Q,\frac{\delta S_T}{\delta Q_R}]\;,
\end{align}
where $S_T$ is the boundary action, see Appendix \ref{sec:GeneralExpression} for derivation,
\begin{align}
    iS_T &= \int_{\partial\Omega} dS \sum_n\text{Tr ln}\Big(\mathbf{1}+i\pi\nu Q_{R}\vec{h}\cdot\vec{\sigma}\tau_{3}\nonumber\\&+(\pi\nu)^2 Q_{R}W_{n}^{\dagger}(\mathbf{1}+i\pi\nu Q_{L}\vec{h}\cdot\vec{\sigma}\tau_{3})^{-1}\tilde{Q}_{L}W_{n}\Big)\; .
\end{align}
The charge current can be calculated from these matrix currents as \begin{align}j^c_{L,R} &= -\frac{e}{2}\text{tr}\rho_1\tau_3\vec{\mathcal{J}}_{L,R}(t,t)\nonumber\\&-\frac{e}{2}\int \frac{dE}{2\pi\hbar}\text{tr}\rho_1\tau_3\vec{\tilde{\mathcal{J}^n}}_{L,R}\label{eq:chargeCurrent}\;,\end{align}
where $\mathcal{J}_{L,R}^n$ is the contribution of the $n$th mode to $\mathcal{J}_{L,R}$. 
In the limit of many channels, the sum over modes can be converted into an integral over the incoming momentum, characterized by their angle $\phi$ with the interface normal, $\mathcal{J}_{L,R}^n\rightarrow \mathcal{J}_{L,R}^{\phi}$, so that
\begin{align}
j^c_{L,R}= -\frac{e N_{\text{ch}}}{4}\int_{-\frac{\pi}{2}}^{\frac{\pi}{2}} \cos\phi d\phi\int \frac{dE}{2\pi\hbar}\text{tr}\rho_1\tau_3\vec{\mathcal{J}}^{\phi}_{L,R}\;,\end{align} where $N_{\text{ch}}$ is the number of conduction channels per spin of the point contact.

The definition of the variational derivative under the normalization constraint, states that for all test matrices $\alpha$ we have
\begin{align}
    iS_{T}(Q_L+[\alpha,Q_L],Q_R)-iS_T(Q_L,Q_R) = \text{tr}\alpha\mathcal{J}_L + O(\alpha^2)\;.
\end{align}
Specifically, for $\alpha = \chi \rho_1 \tau_3$ where $\chi\in\mathbb{R}$ this implies
\begin{align}
    \chi j^c_L &=-\frac{e}{2}\Big(iS_{T}(Q_L+[\chi\rho_1\tau_3,Q_L],Q_R)\nonumber\\&-iS_{T}(Q_L,Q_R)+O(\chi^2)\Big)\;.
\end{align}
Using that the expansion of $e^{-i\frac{\chi}{2}\rho_1\tau_3}Q_Le^{i\frac{\chi}{2}\rho_1\tau_3}$ gives in first order $-\frac{i}{2}[\chi\rho_1\tau_3,Q_L]$, this condition can be written as
\begin{align}
    j^c_L &= 
    \frac{ie}{t_0}\partial_{\chi}S_{\text{FC}}(\chi)|_{\chi = 0}\;,
\end{align}
where $t_0$ is the measuring time and $S_{\text{FC}}(\chi)$ is given by
\begin{align}
    S_{\text{FC}}(\chi)& = -\frac{t_0}{2} S_T(e^{-i\frac{\chi}{2}\rho_1\tau_3}Q_Le^{i\frac{\chi}{2}\rho_1\tau_3},Q_R)\;.
\end{align}
Using cyclic permutation under the trace, and the fact that $W_{n}$ and $\vec{h}\cdot\vec{\sigma}\tau_3$ both commute with $\tau_3$, we find that $S_{\text{FC}}$ satisfies
\begin{align}
    S_{\text{FC}}(\chi)& = S_T(Q_L,e^{-i\frac{\chi}{2}\rho_1\tau_3}Q_Re^{i\frac{\chi}{2}\rho_1\tau_3})\;.
\end{align}
For the right side of the boundary we can make an entirely similar argument, for all $\beta$ we have
\begin{align}
    -iS_{T}(Q_L,Q_R+[\beta,Q_R])+iS_T(Q_L,Q_R) = \text{tr}\beta\mathcal{J}_R + O(\beta^2)\;,
\end{align}
which for $\beta = \frac{\chi}{2} \rho_1\tau_3$ implies
\begin{align}
    j^c_R = \partial_\chi S_{\text{FC}}(\chi)|_{\chi = 0} = j^c_L\;.
\end{align}
This shows that charge current is conserved. Moreover, it shows that $S_{\text{FC}}$ is in fact the full counting statistics action.

An analogous procedure allows us to study thermal currents using $\alpha_E = \chi_{E}E\rho_1$ where $\chi\in\mathbb{R}$. Also $\alpha_E$ commutes with $\vec{h}\cdot\vec{\sigma}\tau_3$ and $W_{n}$, so that we may write a single full counting statistics for thermal currents:
\begin{align}
    S_{\text{FC},E}(\chi_E) = S_T(Q_L, e^{-i\frac{\chi_{E}}{2}E\rho_1}Q_R e^{i\frac{\chi_{E}}{2}E\rho_1})\;,
\end{align}
which upon differentiating gives the thermal current.

For spin currents we need to write two different full counting statistics action for the right and left side of the boundary,
\begin{align}
    S_{\text{FC}\sigma_a,R}(\chi_{S,R}) &= S_T(Q_L, e^{-i\frac{\chi_{S,R}}{2}\rho_1\sigma_a}Q_R e^{i\frac{\chi_{S,R}}{2}\rho_1\sigma_a})\;,\\
    S_{\text{FC}\sigma_a,L}(\chi_{S,L}) &= S_T(e^{i\frac{\chi_{S,L}}{2}\rho_1\sigma_a}Q_Le^{-i\frac{\chi_{S,L}}{2}\rho_1\sigma_a}, Q_R)\;.
\end{align}
Since $\sigma_a$ does not necessarily commute with $\vec{h}\cdot\vec{\sigma}\tau_3$ or $W_{n}$, those two actions are distinct, reflecting the non-conservation of spin currents. Similarly, the spin energy full counting statistics for either side of the boundary, defined with $\alpha_{SEL,R} = \chi_{SEL,R}E\rho_1\tau_3\sigma_a$, are distinct.
If one of the materials is clean, a similar principle can be used to define the full counting statistics for charge, energy, spin and spin-energy currents.

\section{Zero bias conductance peaks}\label{sec:anomalousproximity}
%\SB{I have several comments to this section, which in my view needs revision. In the very beginning one should define what we mean with "anomalous PE". What makes it anomalous?  some other comments in the text}
%In the previous sections, we have laid out the formalism and derived boundary conditions for spin-active interfaces with arbitrary transparency, showing that they can be applied to junctions with dirty materials on both sides as well as junctions with one dirty and one clean material. In the rest of this paper we discuss different applications. 
\begin{figure}
    \centering
    \includegraphics[width=8.6cm]{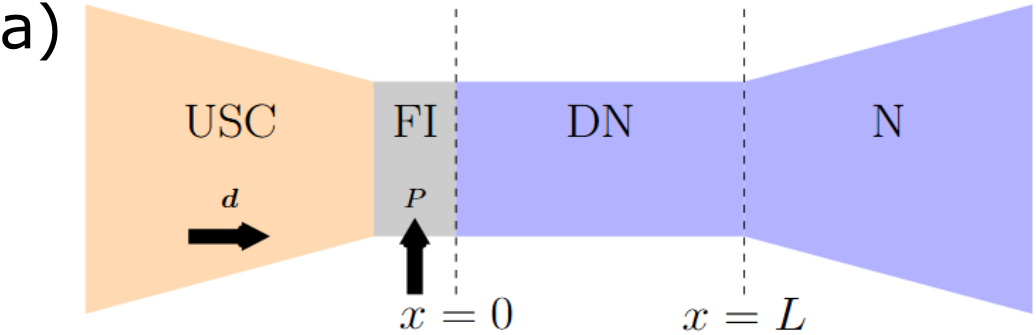}
    \includegraphics[width=6.6cm]{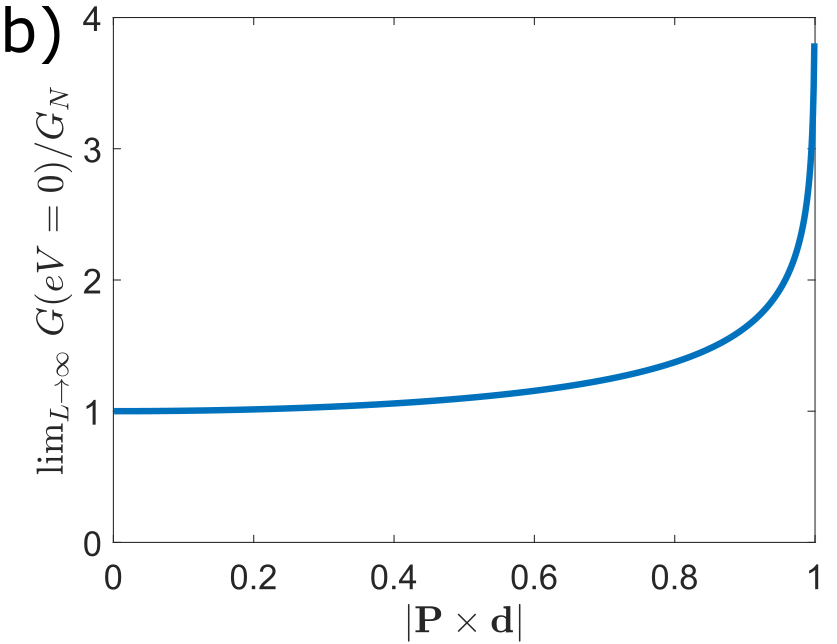}
    \caption{\textbf{a)} The setup used for the anomalous proximity effect. \textbf{b)} The ratio of the zero bias conductance and the normal state conductance for long junctions, $L\gg \frac{\sigma_N}{G_B}$. } 
    \label{fig:AP}
\end{figure}

In this section, we use the derived boundary conditions to describe  the proximity effect of unconventional superconductors on a dirty normal metals when the boundary between them is spin-active. In particular we focus on  the resulting zero bias conductance signatures.
To this end, we consider the junction illustrated in Fig. \ref{fig:AP}(a) in which a diffusive normal bar is sandwiched between \textit{p}-wave superconducting and normal electrodes, with a spin-active barrier towards the superconductor. 

Similar junctions without a spin-active barrier have been studied in detail in literature. The proximity effect of \textit{p}-wave superconductors has several features that differ from that of conventional superconductors, and its characteristic features are therefore termed the anomalous proximity effect.
Examples are the appearance of odd-frequency triplet pairs in the normal region \cite{tanaka2004anomalous,tanaka2005anomalous,tanaka2007theory,tanaka2011symmetry,tanaka2024theory} and the strong proximity effects due surface Andreev bound states (SABS) \cite{buchholtz1981identification,hara1987quasiclassical,hu1994midgap,tanaka1995theory,yokoyama2011anomalous,tanaka2024theory}. For 1D $\textit{p}$-wave or 2D $p_x$-wave superconductors a SABS appears at zero energy (ZESABS) \cite{tanaka2022theory}, leading to a strong zero bias conductance peak which reaches the quantized value $2G_Q$ per channel, where $G_Q$ is the conductance quantum, regardless of the normal state conductivity and length of the bar \cite{tanaka2004anomalous,tanaka2005anomalous,ikegaya2016quantization}. %\SB{The GF is a mathematical object. I would prefer to emphasize the physics. Is the divergence related to surface bound states?  }

Before turning to the results for spin-active junctions, we briefly recap the features in for junctions with spin-independent boundaries. In such junctions, the transparency does not affect the strength of the proximity effect of the ZESABS \cite{tanaka2004anomalous,tanaka2005anomalous,tanaka2005theory,tanaka2022theory}. Indeed, for low energies, the pair potential for the incoming mode is $\hat{\Delta}_{\text{in}} = \Delta \tau_2\vec{d}\cdot\vec{\sigma}$, where $\vec{d}$ is the \textit{d}-vector, while for the outgoing mode $\hat{\Delta}_{\text{out}} = -\Delta\tau_2\vec{d}\cdot\vec{\sigma}$, %\SB{why do we need here indices 1,2? maybe better "in" "out"'?}
so that the boundary Green's function $X$ in Eq. (\ref{eq:magn_X}) reduces to the surface Green's function of the superconductor, which reads \cite{tanaka2004anomalous,tanaka2022theory}:
\begin{align}
    X_n &\approx T_{1n}\frac{\Delta}{E}\tilde{C}\;,\\
    \tilde{C} &= \begin{bmatrix}
        i&\vec{d}\cdot\vec{\sigma}\\\vec{d}\cdot\vec{\sigma}&-i
    \end{bmatrix}\;,
\end{align} %\SB{all this is too technical and  difficult to follow. At least we should may connection with previous equations/results, notations, etc }
Now consider the spin-independent Tanaka-Nazarov boundary condition. 
For energies $|E| \ll T_{1n}\Delta$, the denominator in Eq.~(\ref{eq:magn_IR}) is 
dominated by the term proportional to $X$, and therefore the matrix current reads \cite{tanaka2003circuit,tanaka2004anomalous,tanaka2022theory}
\begin{align}\label{eq:Cnospin}
    I = \{\tilde{C},G\}^{-1}[\Tilde{C},G] + O(E^2)\;,
\end{align}
which is independent of transparency, as found previously \cite{tanaka2005anomalous}.
\subsection{Effect of spin-dependent transmission on the anomalous proximity effect}
To illustrate the effects of a spin-dependent transparency and an interfacial exchange field $\vec{h}$, we consider them separately. Moreover, for illustrative purposes, we will consider a junction with a single channel per spin and thus drop the channel index $n$. Inclusion of multiple channels only affects the results quantitatively, not qualitatively. Let us first consider a fully polarized barrier, that is $t = u$ in Eq. (\ref{eq:Wspinfilter}), which gives a  spin-dependent transparency $(\pi\nu W)^2=T_{1} \frac{1}{2}(1+\tau_3\vec{P}\cdot\vec{\sigma})$, where $\vec{P}$ the polarization direction. In this case, using $C = \frac{1}{E}(i\sqrt{\Delta^2-E^2}\tau_3 + \Delta\tau_2)$ \cite{tanaka2022theory}, we obtain an effective surface Green's function defined as % \SB{it would be useful to refer to a previous  equation }
\begin{align}\label{eq:EffectivesurfaceGreensfunction}
    \tilde{C} = W^{\dagger}CW &\approx \frac{\Delta}{E}\frac{1}{2}(1+\tau_3\vec{P}\cdot\vec{\sigma})\begin{bmatrix}
        i& |\vec{P}\times\vec{d}|\\|\vec{P}\times\vec{d}|&-i
    \end{bmatrix}\;.
\end{align}
Thus, the effective surface Green's function still has a pole at zero energy. For such surface Green's functions the denominator in Eq.~(\ref{eq:magn_IR}) for small energies is still dominated by $X$ and hence the proximity effect at zero energy 
is still independent of transparency, $T_{1}$. However, the magnitude of the pair amplitudes, and thus the conductance, is affected by the spin-filtering in two ways. First of all, due to the projection, only one type of spin can pass through a spin-filter, so that any proximity effect is completely spin-polarized and the maximum attainable conductance is $G_Q$ rather than $2G_Q$. Next to this, the magnitude of the pair amplitudes depends on the angle between $\vec{P}$ and $\vec{d}$. 

If $\vec{P}$ and $\vec{d}$ are orthogonal, all triplets in the unconventional superconductor are equal spin triplets along the $\vec{P}$-axis. Therefore, the effective surface Green's function $\tilde{C}$ for the surviving spin species has the same form as in S/N junctions with a spin-independent barrier, Eq.~(\ref{eq:Cnospin}), in which the pair amplitudes have the same magnitude as the the $\tau_3$ component.
Due to this, its conductance is independent of the normal state conductivity of the metal and quantized at $G = G_Q$. %The 
If $P$ and $d$ are not orthogonal on the other hand, the superconductor a fraction $|\vec{P}\cdot\vec{d}|$ of the triplets in the superconductor have zero projection along the $\vec{P}$-axis, which do not survive the barrier. Therefore, the pair amplitudes in Eq.~(\ref{eq:EffectivesurfaceGreensfunction}) are suppressed, leading to a suppression of the proximity effect. %quantization is lost\SB{I don't think quantization is the right word here and below. }.
To find the resulting pair amplitudes in the normal metal,  we use the $\theta$-parameterization for the spin-projected retarded Green's function in Nambu space,
\begin{align}
    g^R = \begin{bmatrix}
    \cosh\theta&-i\sinh\theta\\-i\sinh\theta&-\cosh\theta
\end{bmatrix}\;.
\end{align}
The boundary condition for the $\theta$ parameter at the spin-active interface with the superconductor reads
\begin{align}
    \frac{\partial \theta}{\partial x}(x = 0) &= \frac{2 G_B}{\tilde{T}\sigma_N}\frac{\sinh\theta-|\vec{P}\times\vec{d}|\cosh\theta}{\cosh\theta-|\vec{P}\times\vec{d}|\sinh\theta}\;,\label{eq:BCtheta}
\end{align}
where $G_B$ is the normal state conductance of the barrier, $\tilde{T}$ its average transparency and $\sigma_N$ the normal state conductivity of the bar.
%\SB{one need to specify that $x=0$ in this expression}
For $|\vec{P}\times\vec{d}| = 1$, this reduces to $\frac{\partial\theta}{\partial x} = -\frac{G_B}{\tilde{T}\sigma_N}$. 

For $E = 0$, the Usadel equation reads $\partial_{xx} \theta = 0$. %\SB{so?} We assume that the contact with the normal metal is very good\SB{the NN contact. We have 2 figures with several structures, why not to show the one analyzed here to make the reading easier?}, as is often the case if it is the same material, 
Since we assume that the contact with the normal metal electrode is good, $\theta(x = L)$ = 0. Thus, 
\begin{align}\label{eq:thetalinear}
    \theta(x) = \theta(x = 0)\frac{L-x}{L}\;,
\end{align}
where $\theta(x = 0)$ is fixed by the boundary condition for the spin-active interface at $x = 0$, Eq. (\ref{eq:BCtheta}). If $\vec{P}$ and $\vec{d}$ are orthogonal, $\theta$ drops out of the right hand side of Eq. (\ref{eq:BCtheta}), and hence
\begin{align}\label{eq:thetaperp}
     \theta(x = 0, |\vec{P}\times\vec{d}| = 1) = \frac{2 G_B L}{\tilde{T}\sigma_N} = \frac{2G_B}{\tilde{T} G_N}\;,
\end{align}
which has been shown to lead to conductance quantization $G(eV = 0) = G_Q$, independent of the conductivity and length of the normal metal bar \cite{tanaka2004anomalous}. %However for $|\vec{P}\times\vec{d}| < 1$, $|\frac{\partial \theta}{\partial x}|<\frac{G_B}{\sigma_N}$, that is, the proximity effect is weakened, and conductance quantization can not anymore be expected. 
In what follows, we show that this conductance quantization is lost if $\vec{P}$ and $\vec{d}$ are orthogonal, and that the zero bias conductance is proportional to the normal state conductance  $G_N = \frac{\sigma_N}{L}$.

%If $\vec{P}$ and $\vec{d}$ are perpendicular, we obtain the same expressions as for a spinless boundary, %\SB{what is a normal boundary}, 
%and therefore quantization of conductance remains\SB{why?}. On the other hand, 
%This is a transcendental equation. If $\vec{P}$ and $\vec{d}$ are parallel, it simplifies we obtain,
To this end, we first consider another limit, $\vec{P}\times\vec{d} = 0$. In this case all triplets have zero projection along $\vec{P}$, and therefore they do not penetrate the barrier, the effective surface Green's function $X$ is diagonal in Nambu space. Due to this, Eqs.~(\ref{eq:thetalinear}, \ref{eq:thetaperp}) simplify to
\begin{align}
    \theta(x = 0, |\vec{P}\times\vec{d}| = 0) = -\frac{2 G_B}{\tilde{T}G_N}\tanh\theta(x = 0)\;,
\end{align}
which is solved by $\theta (x = 0, |\vec{P}\times\vec{d}| = 0) = 0$, that is, the proximity effect is absent. 
%\SB{for  the minus sign in 58 is crucial. I assume it is checked:) } \TK{The minus sign is actually not that crucial, $\theta = 0$ is always a solution. But also, yes it is checked :)}\SB{but if there is no  minus otehr solutions might be possible? } \TK{True, if $\frac{G_B}{G_N}$ is in the right range. You could thus even view it as that the minus sign must have this sign because otherwise there would be spurious solutions (and $\theta(x = 0)$ is the physical one because the surface Green's function is diagonal in Nambu space in this limit.)} 

For any other orientation, $0<|\vec{P}\times\vec{d}|<1$, and therefore one obtains a transcendental equation from Eqs.~(\ref{eq:thetalinear}, \ref{eq:thetaperp}):
\begin{align}\label{eq:retaredGeneral}
    \theta(x = 0) = \frac{2 G_B L}{\tilde{T}\sigma_N} \frac{\tanh\theta(x = 0)-|\vec{P}\times\vec{d}|}{1-|\vec{P}\times\vec{d}|\tanh\theta(x = 0)} \;.
\end{align}

For large $L\gg \frac{\sigma_N}{G_B}$, the right hand side of Eq.~(\ref{eq:retaredGeneral}) diverges unless
$\tanh\theta \approx |\vec{P}\times\vec{d}|$, which implies that for long junctions the solution satisfies
\begin{align}\label{eq:thetanotparorperp}
    \theta(x = 0, 0<|\vec{P}\times\vec{d}|<1) \approx \tanh^{-1}|\vec{P}\times\vec{d}|\;.
\end{align}
For any orientation such that $0<|\vec{P}\times\vec{d}|<1$ this converges to a nonzero and finite value that does not grow indefinitely with $L$. This is in great contrast with the perpendicular orientation and the spin-independent barrier, both of which obey Eq.~(\ref{eq:thetaperp}). %Thus, for any misalignment of $\vec{P}$ and $\vec{d}$, $\theta (x = 0)$ converges to a finite value as $L\xrightarrow{}\infty$, thereby showing distinctly different behaviour compared to the spinless interface.%\SB{But 57 is also finite in the spinless case.}\TK{1) 57 does not contain a spinless case, so not entirely sure what you mean by that. 2) In the spinless case, and also if $|\vec{P}\times\vec{d}| = 1$, this actually becomes infinite as $L\xrightarrow{}\infty$, because $G_N$ diverges. To reflect this more clearly, I will write it in terms of $\sigma_N$ and $L$}
Since the zero bias conductance is directly determined by $\theta(x = 0)$ \cite{tanaka2003circuit,tanaka2004anomalous}, this implies that the zero bias conductance deviates from $G_Q$. Specifically, in the long junction limit where $G_B\gg G_N$, we obtain
\begin{align}
    \frac{G(eV = 0)}{G_N} &= \Bigg(\frac{1}{L}\int_{0}^{L}dx \frac{1}{\cosh^2 (\tanh^{-1}|\vec{P}\times\vec{d}|)\frac{x}{L}}\Bigg)^{-1}\nonumber\\&= \frac{\tanh^{-1}|\vec{P}\times\vec{d}|}{|\vec{P}\times\vec{d}|}\;.
\end{align}
Thus, unless $\vec{P}$ and $\vec{d}$ are perpendicular, the zero bias conductance in a junction with a spin-polarized interface is not quantized to $G_Q$, but instead scales with the normal state conductance, and thus decreases with the length of the junction. The proportionality constant is shown in Fig. \ref{fig:AP}(b). It approaches 1 as $|\vec{P}\times\vec{d}|\xrightarrow{}0$, reflecting the absence of equal spin triplets and thus of a proximity effect if the two vectors are parallel, while it diverges as $|\vec{P}\times\vec{d}| \xrightarrow{}1$, reflecting that $G_Q / G_N$ diverges in the long junction limit.
%Unless $\vec{P}$ and $\vec{d}$ are perpendicular, this is finite. \SB{I do not see that. Probably I am missing an important point or definition of what you mean with quantization. It would be nice to explain it in the introduction together with the concept of anomalous PE}
%Therefore there is no conductance quantization, and the conductance of the junction scales with $G_N$.

%Also for intermediate spin-polarization, the proximity effect is transparency independent. In this case, the effective $\textit{d}$-vector of  $W^{\dagger}CW$ attains both a component along $\vec{d}$ and one along $\vec{P}$.

\subsection{Effect of the interfacial exchange field on the anomalous proximity effect}
Next we consider the effect of a boundary exchange field on the anomalous proximity effect. For simplicity of presentation we consider the absence of any polarization and again, we consider a single channel so that we may ignore the channel index. In this case, in the quantity $X$ defined in Eq.~(\ref{eq:magn_X}) we have to set $Q_L \rightarrow C$ and $\pi \nu W\xrightarrow{}\sqrt{T_{1}}$, so that
\begin{align}\label{eq:XC}
    X &=  i\pi\nu \vec{h}\cdot\vec{\sigma}\tau_3 + T_{1}(1 + i\pi\nu C\vec{h}\cdot\vec{\sigma}\tau_3)^{-1}C\;.
\end{align}
For energies $|E| \ll \Delta, h$, we have,
if $\vec{h}$ is perpendicular to $\vec{d}$, as shown in Appendix \ref{sec:XfromCexchangefield},
\begin{align}\label{eq:Xperp}
    X \approx \frac{T_{1}}{1+(\pi\nu h)^2}C\;, 
\end{align}
that is, there is still a pole at $E = 0$, the surface Andreev bound state is not affected and the zero bias conductance remains quantized at $2\frac{e^2}{h}$. However, the peak width changes. The peak width $E_w$ is defined as the energy for which $X$ starts dominating the denominator, i.e. $\text{tr}((\tau_1-i\tau_2)X(E_w))\sim 1$. Thus, the peak width is given by $E_w \sim \frac{T_{1}\Delta}{ (1+(\pi\nu h)^2)}$, a factor $1+(\pi\nu h)^2$ smaller than without the magnetic interface. We conclude that a perpendicular exchange field does not change the height of the zero bias conductance peak, but makes it sharper.%\SB{exchange field makes teh peak sharper?}\TK{Yes, wrote that explicitly now}

On the other hand, if $\vec{h}$ is parallel to $\vec{d}$, we obtain, see Appendix \ref{sec:XfromCexchangefield} for details,
\begin{align}\label{eq:Xparallel}
    X \approx i\pi\nu(1-\frac{T_{1}}{(\pi\nu h)^2})\vec{h}\cdot\vec{\sigma}\tau_3 + \frac{T_{1}}{\pi\nu h}\tau_1\;,
\end{align}
Thus, there is no divergence at $E = 0$. This means the effective boundary Green's function loses its zero energy surface Andreev bound state, and like discussed in the previous section, this means the conductance is not quantized, but scales with $G_N$ in the long junction limit. Moreover, the introduced pairs at zero energy are purely singlet. This is the counter part of the a similar effect in singlet superconductors, for which a spin-active interface can result in a proximity effect that is completely triplet at $E = 0$ \cite{linder2010signature}.  This can be used as a test for the pairing symmetry of triplet superconductors. 
\section{Josephson junctions with spin-active interfaces}\label{sec:JJ}
\begin{comment}
\subsection{0-$\pi$ transition in BCS superconductors}
The first example we may consider is the junction between two BCS superconductors. First, we assume that the transparency of the junction is small and does not depend on spin. In that case the electrical current between the two is
\begin{align}
    I = \sum_{\omega}\frac{8\sin\phi(1-(\pi\nu h)^{2})\Delta^{2}}{(1-(\pi\nu h)^{2})^{2}\Delta^{2}+(1+(\pi\nu h)^{2})^{2}\omega_n^{2}} = \frac{2\pi\Delta}{1+(\pi\nu h)^{2}}\tanh{\Big(\frac{\pi}{2}\frac{\Delta}{T}\Big)}\frac{1-(\pi\nu h)^{2}}{1+(\pi\nu h)^{2}}\sin\phi.
\end{align}
As expected, in the electrical current there are no terms of odd order in $h$.
For $\pi\nu h = 1$ there is a sign reversal of the current, while for $h\gg \Delta$ the current is severely suppressed. If neither of the two BCS superconductors has a spin-accumulation, no spin current is induced either.
\end{comment}
In this section, we apply the boundary condition Eq.~(\ref{eq:magn_IR}) to two cases: a junction between two spin-split superconductors, and a junction between a spin-split superconductor and a $p$-wave superconductor. Because we are studying non-dissipative transport, we use in this section  the Matsubara formalism, and convert the energy integrals to a sum over a discrete set of Matsubara frequencies $\omega_n = (2n+1)\pi T$ on the imaginary energy axis, that is, in our boundary condtions we set $E\rightarrow i\omega_n$.

\subsection{Junction between spin-split superconductors}
\begin{figure}
    \centering
    \includegraphics[width=6.6cm]{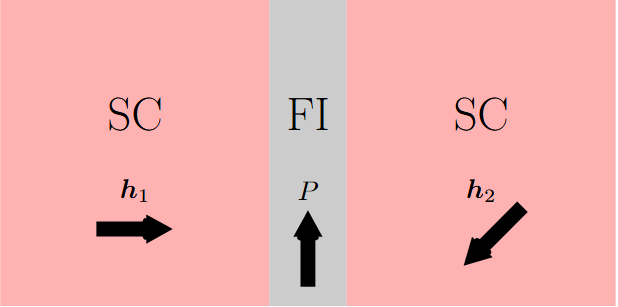}
    \caption{Schematic overview of the junction with two spin split superconductors seperated by a spin-active barrier.}
    \label{fig:SpinSplitJunction}
\end{figure}

We consider a junction between two dirty spin-split superconductors, separated by a boundary with a spin polarization. We assume the three magnetizations at the boundary and left and right materials are mutually orthogonal, in which case non-reciprocal effects such as $\phi_0$-effects and diode effects may be expected \cite{silaev2017anomalous,schulz2025quantum}. A sketch of the junction is shown in Fig.~\ref{fig:SpinSplitJunction}. This situation has been analyzed in Ref.~\cite{silaev2015spin}, but only in the strict tunnelling limit. Here we consider also the effect of terms higher order in transparencies, which allows us to discuss the superconducting diode effect. %\SB{Are we going beyond the tunneling limit?}\TK{In principle the eigenvalues of $W$ can become 1}

%\SB{We shoudl teh notation of teh previous section, where ther is no $C$ but $Q$'s right?}
%Using Eq.~(\ref{eq:Cexpression}), we have in Matsubara representation \SB{from eq. 25 it is difficult to write this expression. Moreover a sentence or two  about how to go from real frequency to Matsubara would also be Ok, or at least writing, in The matsubara representation .....}
Because dirty spin-split superconductors are isotropic, we can use $Q$-matrices for both sides. If both materials can be treated as reservoirs, we can use their bulk quasiclassical Green's functions,  
\begin{align}
    Q_L & = C = \sum_{\pm}\frac{1}{\sqrt{(\omega_n\pm ih)^{2}+\Delta^{2}}}\Big((\omega_n\pm ih)\tau_{3}\nonumber\\&+\Delta e^{i\phi/2\tau_{3}}\tau_{2}\Big)\otimes (1\pm \cos\phi_h \sigma_x + \sin\phi_h \sigma_{y})/2\;,\\
    Q_R  &= \sum_{\pm}\frac{1}{\sqrt{(\omega_n\pm ih)^{2}+\Delta^{2}}}\Big((\omega_n\pm ih)\tau_{3}\nonumber\\&+\Delta e^{i\phi/2\tau_{3}}\tau_{2}\Big)\otimes (1\pm \sigma_x)/2\;,\label{eq:QJJ}\\
    W &= \sum_{\pm} t_{\pm}(1\pm\tau_{3}\sigma_{z})/2\;,\label{eq:Wtwotransparencies}\\
    \vec{h} &= h_{z}\vec{\hat{z}}\;.
\end{align}
%First we consider the case $h_{z} = 0$. In that case we obtain
%\begin{align}
    %X&=  (\pi\nu)^{2}(t_{+}+t_{-})^{2}C+(t_{+}-t_{-})^{2}\sigma_{z}C\sigma_{z}\nonumber\\&+(t_{+}+t_{-})(t_{+}-t_{-})\{C,\sigma_{z}\}\;, \\
    %\Tilde{X}& =   (\pi\nu)^{2}(t_{+}+t_{-})^{2}C+(t_{+}-t_{-})^{2}\sigma_{z}Q\sigma_{z}\nonumber\\&+(t_{+}+t_{-})(t_{+}-t_{-})\{Q,\sigma_{z}\}\;. 
%\end{align}
%If $t_{+} = t_{-}$, i.e. the tunneling is spin-independent, we recover the usual Nazarov boundary condition, with $X = (2\pi\nu t_{+})^{2}C$. In that case there is no $\phi_{0}$-effect.
%
%If $t_{+} = -t_{-}$, $X = (2\pi\nu t_{+})^{2}\tau_{3}\sigma_{z}C\tau_{3}\sigma_{z}$, that is, the spin-polarization of the other material is effectively negated. Consequently, the results are trivially obtained as

Here $\phi_h$ is the angle between the two magnetizations. An interesting case is that of a perfect spin filter, in which the minority spin can not penetrate through the barrier, that is, $t_{-} = 0$. In that case
\begin{align}
    X_n &= (\pi\nu t_{+})^{2}(C+\tau_{3}\sigma_{z}C\tau_{3}\sigma_{z}+\{C,\tau_{3}\sigma_{z}\}) \nonumber\\&= 2(\pi\nu t_{+})^{2}\Big(\text{Re}\Big(\frac{\omega_n+ih}{\sqrt{(\omega_n+ih)^{2}+\Delta^{2}}}\tau_{3}\nonumber\\&+\text{Im}\frac{\Delta }{\sqrt{(\omega_n+ih)^{2}+\Delta^2}} e^{i\phi_h}\tau_{2}\Big)\otimes (1+\sigma_{z}\tau_{3})
    \Bigg)\;.
\end{align}
The pair amplitudes of $X$ have a $\phi_h$ different phase compared to the pair amplitudes of $C$, which leads to a phase shift in the current phase relation. This is expected, since anomalous currents often appear if $\vec{P}\cdot(\vec{h}_1\times\vec{h}_2)\neq 0$ \cite{halterman2022supercurrent}.
%Next to this, we remark that since triplets on both side are required for any tunneling process to appear, the current shown above is $O(h^{2})$.
%Moreover, there are also diode effects. However, these do not appear for a perfect spin filter because the current takes the form
For the perfect spin-filter, the current phase relation takes the form
\begin{comment}
Indeed, the electrical current reads
\begin{align}
&    \Bigg((h^{4}+\omega^{4})(1+4t_{+}^{2}+8t_{+}^{4})+(1+8t_{+}^{4})\Delta^{4}+4t_{+}^{2}(1+2t_{+}^{2})|\Delta^{2}+(\omega_n+ih)^{2}|\omega_n^{2}-8t_{+}^{4}|\Delta^{2}+(\omega_n+ih)^{2}|+2(1+2t_{+}^{2})\omega_n^{2}\Delta^{2}\nonumber\\&-2(1+2t_{+}^{2}+8t_{+}^{4})\Delta^{2}h^{2}+2(1+2t_{+}^{2}+4t_{+}^{4})\omega_n^{2}h^{2}+2t_{+}^{2}(1+2t_{+}^{2})|\Delta^{2}+(\omega_n+ih)^{2}|+4t_{+}^{2}\Delta^{2}(\omega_n^{2}+\Delta^{2}-h^{2}\nonumber\\&-|\Delta^{2}+(\omega_n+ih)^{2}|)\sin\phi\Bigg)^{-1}\nonumber\\&4\Delta^{2}t_{+}^{2}\text{Im}\sqrt{(\omega_n+ih)^{2}+\Delta^{2}}\cos\phi
\end{align}
\end{comment}
\begin{align}
    I = \sum_{n}\frac{A_{1}(\omega_n)\cos\phi}{B_{1}(\omega_n)+C_{1}(\omega_n)\sin\phi}\;,
\end{align}
where $A_{1},B_{1},C_{1}$ are lengthy expressions that can be found in the supplementary Mathematica file. %\textcolor{red}{(Mathematica file Spinfiltering\_Spin\_Split_Overleaf.nb)}
We note that the appearance of $\cos\phi$ leads to a maximal $\phi_0$ effect, $\phi_0 = \frac{\pi}{2}$. However, there are no diode effects since $I(\phi) = - I(\pi-\phi)$, that is, the current phase relation is antisymmetric around $\phi = \frac{\pi}{2}$. Thus, for perfect spin-filters there is no diode effect, only a phase shift. 

%Indeed, for the perfect filter we may project on a single spin and $X$ reads
%\begin{align}
%    X &=2(\pi\nu t_{+})^{2}(1+\sigma_{z})/2\otimes\Big(\text{Re}\frac{\omega_n+ih}{\sqrt{(\omega + ih)^{2}+\Delta^{2}}}\tau_{3}\nonumber\\&+\big(e^{i\phi_{h}\tau_{3}}\tau_{1}\text{Im}\frac{\Delta}{\sqrt{(\omega_n+ih)^{2}+\Delta^{2}}}\Big)\\
 %   \phi_{h} &= \phi+\arctan \frac{h_{x}}{h_{y}}\;.
%\end{align}
\begin{comment}
\begin{align}
    X &= 2(\pi\nu t_{+})^{2}(1+\sigma_{z})/2\Big(\sum_{\pm}\frac{1}{\sqrt{(\omega\pm ih)^{2}+\Delta^{2}}}\begin{bmatrix}
        \omega\pm ih \Delta e^{i\phi/2}\\\Delta e^{-i\phi/2}&-(\omega\pm ih)
    \end{bmatrix}\otimes (1+\vec{h}\cdot\vec{\sigma})\Big)(1+\sigma_{z})/2\nonumber\\&=2(\pi\nu t_{+})^{2}\begin{bmatrix}
        \text{Re}\Big(\frac{\omega_n+ih}{\sqrt{(\omega_n+ih)^{2}+\Delta^{2}}}\Big)&(ih_{x}+h_{y})e^{i\phi/2}\text{Im}\frac{\Delta }{\sqrt{(\omega_n+ih)^{2}}}\\(-ih_{x}+h_{y})e^{-i\phi/2}\text{Im}\frac{\Delta }{\sqrt{(\omega_n+ih)^{2}}}&-\text{Re}\Big(\frac{\omega_n+ih}{\sqrt{(\omega_n+ih)^{2}+\Delta^{2}}}\Big)
    \end{bmatrix}\otimes (1+\sigma_{z}\tau_{3}) \;,
\end{align}
\end{comment}
%which shows that $\vec{h}$ enters exactly as a phase $\arctan \frac{h_{x}}{h_{y}}$ in the case of a perfect spin filter.

\begin{figure}
    \centering
    \includegraphics[width=8.6cm]{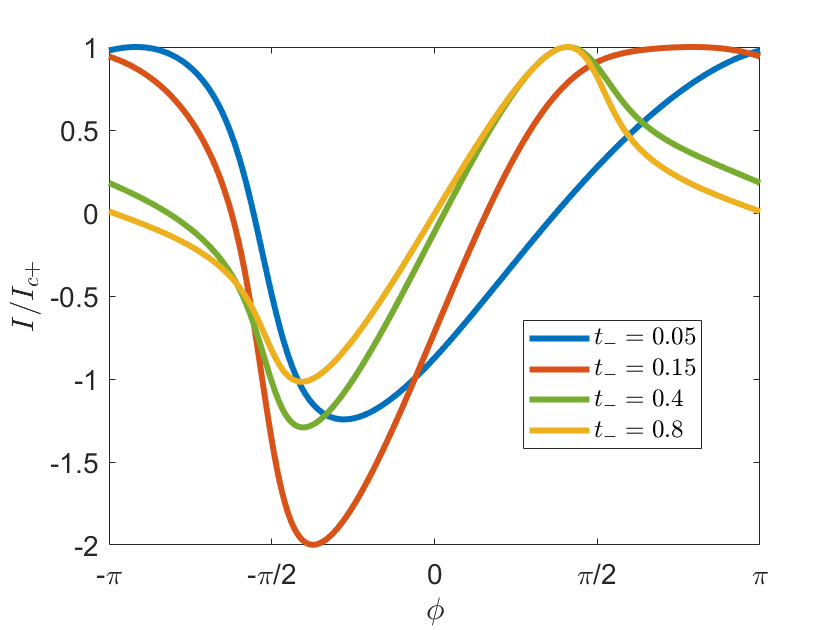}
    \caption{The current phase relation in a junction between two spin-split superconductors with a spin-active boundary condition with $t_{+} = 1,~h = \Delta(T,h),~T = 0.01\Delta(T,h)$ and varying $t_{-}$. For $t_{-} = 0.15$, the current phase relation satisfies $|I_{-}| = 2I_{+}$, indicating a large diode efficiency $\eta\approx\frac{1}{3}$.}
    \label{fig:EtaBarrier}
\end{figure}

For non-zero and non-equal values of $t_{\pm}$, we obtain diode effects, 
as shown in Fig.~\ref{fig:EtaBarrier}. This superconducting diode effect requires the dominant spin transparency to be large in order to have a nonsinusoidal current phase relation. The subdominant spin transparency should be nonzero, but much smaller than the dominant spin transparency. %\textcolor{red}{Should we investigate the parameter dependence? On the one hand it might be good to be thorough, on the other hand, the derivation of the B.C.s is the main result, not the (relatively weak) superconducting diode effect which seems to obey similar dependence as most other diode effects (better at low temperatures, need exchange fields, but not too large because they destroy the superconductivity etc.)}

\subsection{Junction between a spin-split superconductor and a \textit{p}-wave superconductor}
\begin{figure}
    \centering
    \includegraphics[width=6.6cm]{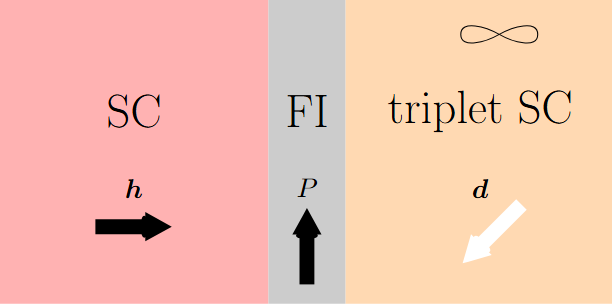}
    \caption{Schematic overview of the junction between a spin split superconductors and a \textit{p}-wave superconductor separated by a spin-active barrier. Magnetizations in the barrier have been indicated using black arrows, the \textit{d}-vector is shown in white to emphasize the different nature of this vector compared to the magnetizations. The triplet superconductor is assumed to be clean so that unconventional superconductivity can persist, the spin-split superconductor is assumed to be dirty.}
    \label{fig:TripletJunction}
\end{figure}

Now consider a junction between a spin-split superconductor and a \textit{p}-wave superconductor in which the boundary contains magnetization, schematically shown in Fig.~\ref{fig:TripletJunction}. For simplicity of presentation, we consider a 1D \textit{p}-wave superconductor. In this case, since time-reversal reverses the magnetization in both the spin-split superconductor and the boundary, but not the \textit{d}-vector in the \textit{p}-wave superconductors, that is there does not exist any diode effect proportional to $\vec{d}\cdot (\vec{h}\cross\vec{P})$, since this object is even under time-reversal.

%The absence of this effect can be shown formally as follows. Consider the case in which the three (pseudo)-vectors are mutually orthogonal. Combination of the application of time-reversal, followed by a mirror operation in spin space around the d-axis ($g\xrightarrow{}\vec{d}\cdot\sigma g \vec{d}\cdot\sigma$), is equivalent to the single setting the phase $\phi\xrightarrow{}-\phi$. Since under the first operation $I\xrightarrow{}-I$, while the second operation does not affect $I$ we conclude $I(\phi) = -I(-\phi)$.

In the limit in which the minority spin can not penetrate the barrier at all, i.e. $t_{-} = 0$ in Eq.~(\ref{eq:Wtwotransparencies}), the current can be calculated analytically. The surface Green's function in the \textit{p}-wave superconductor is \cite{tanaka2022theory}
\begin{align}
    C = \sum_{\pm} \frac{1}{\omega_n}\begin{bmatrix}
        \sqrt{\omega_n^{2}+\Delta^{2}}&\pm i\Delta e^{i\phi/2}\\\pm i\Delta e^{-i\phi/2}&-\sqrt{\omega_n^{2}+\Delta^{2}}
    \end{bmatrix}\otimes(1\pm\sigma_{y})/2\;,\label{eq:Cpwave}
\end{align}
while for the spin-split superconductor we can still use Eq. (\ref{eq:QJJ}).
%Thus, there are only triplet pair amplitudes within the \textit{p}-wave superconductor, indicating that a nonzero exchange field in the spin-split superconductor is required to have a Josephson effect.

Replacing Eq.~(\ref{eq:QJJ}) by this expression and solving for the resulting current, we obtain a first order coupling between the triplet superconductor and the spin-split superconductor due to the spin-mixing at the interface \cite{asano2003josephson}, 
\begin{align}
    I(\phi)  = \frac{A_{2}(\omega_n)\sin \phi}{B_{2}(\omega_n)+C_{2}(\omega_n)\cos{\phi}}\;.
\end{align}
where $A_2,B_2,C_2$ are given in the supplementary Mathematica file. %(PwaveSpinSplitJunction\_Overleaf.nb). 
This expression indeed satisfies $I(\phi) = -I(-\phi)$, and therefore does not have an anomalous current. Connected to this, even for $t_{-}\neq 0$ the junction does not develop a diode effect if $\vec{d},\vec{h},\vec{P}$ are mutually orthogonal. Indeed, for this orientation, time-reversal followed by a $\pi$-rotation in space around the \textit{d}-vector, implies $I(\vec{\phi},\vec{h},\vec{P},\vec{d}) = -I(-\phi,-\vec{h},-\vec{P},\vec{d}) =  -I(-\phi,\vec{h},\vec{P},\vec{d})$, forbidding anomalous currents and diode effects.

%Indeed, for this orientation, the charge current density, i.e. the $\tau_3$ component of Eq. (\ref{eq:magn_IR}), is odd in the phase difference because the Q-matrix $Q_R$ of the spin-split superconductor, Eq. (\ref{eq:QJJ}), and the surface Green's function $C$ of the \textit{p}-wave superconductor, Eq. (\ref{eq:Cpwave}), both satisfy the symmetry $g(\omega_n,\phi) = \tau_2\vec{d}\cdot\vec{\sigma}g(-\omega_n,-\phi)\vec{d}\cdot\vec{\sigma}\tau_2$, while also the transmission matrix satisfies $W = \tau_2\vec{d}\cdot\vec{\sigma}W\vec{d}\cdot\vec{\sigma}\tau_2$.}
%There are a few more notes we can make about this expression. First of all, $A$ is $O(h)$, due to the requirement that triplets are required for transmission through the barrier. On the \textit{p}-wave side triplets are there automatically. However, the only components with \textit{d}-vector in the x or y direction survives, which means that the z-component vanishes.

%This however, does not mean that diode-effects are completely absent in this type of system.
%While the mutually perpendicular orientation does not give any superconducting diode effect, other orientations may result in superconducting diode effect.
%Indeed, in the discussion above, we have %assumed that the three-vectors are mutually orthogonal, which in the case of only spin-split superconductors is the optimal orientation. However, if they are not mutually orthogonal, 
However, even though there is no diode effect in the mutually perpendicular orientation, the junction between a spin-split superconductor can still harbour superconductor diode effects in other orientations. Indeed, we may construct several higher order time-reversal odd scalars using the d-vector, polarization vector and exchange field such as $\vec{d}\cdot \vec{h}\times(\vec{h}\times \vec{P})$. %or $\vec{d}\cdot \vec{P}\times(\vec{h}\times \vec{P})$, or $\vec{P}\cdot \vec{h}\times(\vec{h}\times \vec{d})$. 
%\SB{this paragraph is not clear enough. Teh absence of diode is because we assumed $t_-=0$? then it has to be said}
For this reason,  the diode effect is allowed in orientations in which the three vectors are not orthogonal, again if $t_{\pm}$ are both nonzero but not equal. As we show below, it is in fact nonzero. % \SB{can we make a stronger statement? or only "may"? or is what you hsow in the next paragraph? it is not clear}

\begin{figure}
    \centering
    \includegraphics[width=8.6cm]{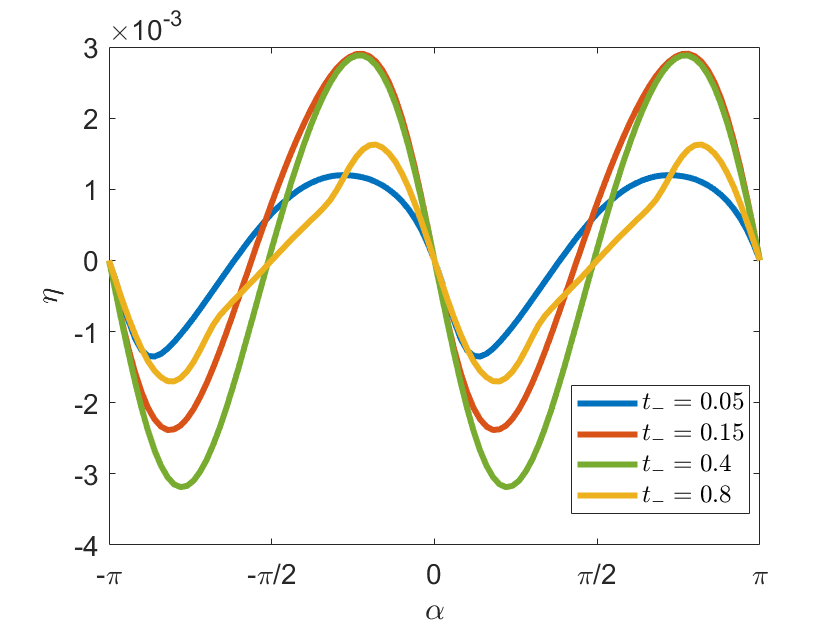}
    \caption{The diode efficiency $\eta$ as a function of the angle between the \textit{d}-vector of the \textit{p}-wave superconductor in the xy-plane if the polarization is along $\frac{1}{\sqrt{3}}(1,1,1)$. We assume the magnitude of the superconducting gaps is the same, $T = 0.01\Delta(T,h)$. Moreover, we set $t_+ = 1$ and $h = \Delta(T,h)$. The polarization of the boundary is chosen such that it is not orthogonal neither parallel to either vector. }
    \label{fig:etapwave}
\end{figure}
%Thus, the restrictions of $\vec{h},\vec{P},\vec{d}$ to obtain a superconducting diode effect are quite stringent, they must not be coplanar and not be mutually perpendicular. If those requirements are met, there are no symmetry restrictions that forbid the superconducting diode effect.
To show that superconducting diode effects can indeed exist in junctions consisting of a triplet superconductor and a spin-split superconductor coupled through a spin-active interface, we fixed the orientation of the polarization vector to $\vec{P} = \frac{1}{\sqrt{3}}(1,1,1)$, the exchange field to $\vec{h} = (1,0,0)$, while varying the orientation of the d-vector as $\vec{d} = (\cos\alpha,\sin\alpha,0)$, so that orientations are achieved in which the three vector are neither coplanar, nor orthogonal. The results are shown in Fig.~\ref{fig:etapwave}. The diode efficiency vanishes in the high symmetry orientations, but the diode efficiency $\eta$ can be nonzero otherwise. Depending on the orientation of the d-vector the diode efficiency may be positive or negative. The diode efficiency remains much smaller than for the junction with spin-split superconductor, reflecting that non-reciprocal effects in spin-split superconductor / FI / triplet superconductor junction require higher order invariants.
%Thus, $\vec{P}$ can not be in the plane spanned by $\vec{h},\vec{d}$, which will be called the xy-plane, nor can it be perpendicular to it. This leaves the option it is in between. From the last symmetry above, it has to satisfy $\eta(\vec{P}_{\parallel},P_{z},\vec{h},\vec{d}) = -\eta(\vec{P}_{\parallel},-P_{z},\vec{h},\vec{d})$. Moreover, $\eta$ should be even in $\vec{d}$ ($I(\vec{d},\phi) = I(-\vec{d},\phi+\pi)$) and odd under the simultaneous sign change of $\vec{h},\vec{P}$ (time-reversal). Spin-mirroring using $\vec{h}\cross\vec{P}$ reverses $\vec{h},\vec{P}$, and the component of $\vec{d}$ in the plane spanned by $\vec{h},\vec{P}$, while keeping $\eta$ the same, which shows that $\eta$ is odd in the component of $\vec{d}$ in the $\vec{h},\vec{P}$-plane. Evenness in $\vec{d}$ implies it is also odd in the component of $\vec{d}$ perpendicular to both $\vec{h},\vec{P}$. \textcolor{red}{The expressions including a finite $t_{-}$ become excessively complicated, I can not really make sense of it.}
\section{Discussion}
Starting from the nonlinear sigma model, we have derived tractable boundary conditions for a spin-active barrier with arbitrary transparency between two materials of which at least one is diffusive. We have shown that in the low-transparency limit our boundary conditions reduce to known results for spin-active interfaces \cite{cottet2009spin,bergeret2012spin}. Moreover, in the limit of spin-independent barriers our expressions agree with the Nazarov and Tanaka-Nazarov boundary conditions \cite{nazarov1999novel,tanaka2003circuit} for any transparency. 

Our boundary conditions therefore provide a generalization of earlier expressions to generic spin-dependent transparencies. The derivation using the nonlinear sigma models allows to obtain compact expressions in all cases. Moreover, since it additionally provides charge, thermal and spin full counting statistics action for the boundary, it allows for the study of effects beyond the stationary boundary conditions studied in this work, such as noise in charge currents, spin currents and thermal currents.

Using the boundary conditions we have studied the anomalous proximity effect of a \textit{p}-wave superconductor through a spin-active interface and shown how the spin-dependent transmission and the  interfacial alter the proximity effect, and how this affects the zero bias conductance.

Moreover, we have studied anomalous Josephson current and diode effect in two types of junctions, S/FI/N junctions and S/FI/S junctions. For the former type of junction, we have shown that both a finite polarization and a finite exchange interaction of a ferromagnetic insulator break the quantization of zero bias conductance in the presence of zero energy Andreev bound states. 

For Josephson junctions with spin-split superconductors, our results show that anomalous Josephson currents appear for spin-active barriers if the magnetizations on the two sides and the one in the barrier together span the whole spin space, according with the constraint $\vec{P}\cdot(\vec{h}_{1}\times\vec{h}_{2})\neq 0$. The anomalous Josephson current is maximized if the three vectors are mutually perpendicular. This anomalous Josephson current is accompanied by a diode effect if the barrier has nonzero filtering but is not a perfect filter, that is, if the minority spin has a non-zero transmission probability that is different from that for the majority spin.  % This indicates that diode effects in this type of junction are more rare and usually weaker than anomalous Josephson currents or $\phi_{0}$-effects.

Next to this, we have studied the junction consisting of a ferromagnetic insulator between a triplet superconductor and a spin-split superconductor. For such junctions the mutually orthogonal orientation does not provide any anomalous Josephson currents, because $\vec{P}\cdot(\vec{h}\times\vec{d})$ is a time-reversal even quantity and thus can not correspond to a time-reversal odd equilibrium current. If the relative orientation between $\vec{h}$ and $\vec{d}$ is not perpendicular anomalous Josephson currents are allowed, corresponding to higher order combinations of $\vec{h},\vec{d}$ and $\vec{P}$. Also for this junction diode effects are only allowed if the barrier is not a perfect filter. However, since the non-reciprocal effect is due to a higher order invariant, diode effects in this type of junction are generally much smaller than in the junction with spin-split superconductors.

%\SB{I am not sure we need materials in this paper. But we may keep this paragraph. What we should emphasize also is the fact that the method we propose for the derivation of BCs can be generalized for other type of materials, magnetic orders etc.  }
To verify the predictions made for the anomalous proximity effect and the Josephson diode effect in junctions with spin-active interfaces we recommend to use spin-split superconductors in which the direction of magnetization is tunable, such as hybrids of EuS or EuO with a standard conventional superconductor, such as Al or Nb. The spin-active boundary can also be realized using EuS or EuO. Candidate materials for triplet superconductors are the uranium based superconductors such as UTe$_2$, in which the superconductivity appears while the material is close to a ferromagnet transition \cite{ran2019nearly,aoki2019unconventional,aoki2022unconventional,kanasugi2022anapole}, and some superconducting states in few layer systems, for which indications of an unconventional gap structure or coexistence of superconductivity and time-reversal symmetry breaking have been found \cite{pantaleon2023superconductivity,han2025signatures,xu2025signatures}.

Lastly, we highlight that while in this paper we have focussed on junctions with spin-split superconductors, unconventional superconductors and spin-active interfaces with a net polarization, the proposed method is not restricted to these type of junctions, but can be applied to a large range of junction types, including those that on the clean side have any type of unconventional magnetism, valley degree of freedom or multiple bands, via the appropriate modification of the surface Green's function and transmission matrices $W, W^{\dagger}$.

\section{Acknowledgements}
We thank Tero T. Heikkilä for useful discussions.  T.K. is supported by the Research Council of Finland through DYNCOR, Project Number 354735 and through the Finnish Quantum Flagship, Project Number 359240. His work is part of the Finnish Centre of Excellence in Quantum Materials (QMAT).   Y. T. acknowledges support from JSPS with Grants-in-Aid for Scientific Research (KAKENHI Grant Nos. 23K17668, 24K00583, 24K00578, 24K00556, 25H00609, 25K07203, and 25H00613). F.~S.~B. thanks financial support from the the Spanish MCIN/AEI/10.13039/501100011033 
through the grant PID2023-148225NB-C31,  the
European Union’s Horizon Europe research and innovation program under grant agreement No.~101130224 (JOSEPHINE), and InstituteQ under the Chair of Excellence programme.

\bibliography{biblio}
\appendix
\newpage\onecolumngrid\
\section{Derivation of general expression}\label{sec:GeneralExpression}
In this Appendix we derive a compact expression for the boundary condition for boundaries with exchange field and spin-dependent transmission. In other words, we derive  Eq.~(\ref{eq:magn_X}) from the action in Eq.~(\ref{eq:general_BC}). 

We may decompose the matrix appearing in the action into lower block diagonal, upper block diagonal and block diagional matrices in two ways:
\begin{align}
    S_T =&\text{Tr ln}\Bigg(\begin{bmatrix}
         \mathbf{1}+G_{L}\vec{h}\cdot\vec{\sigma}\tau_{3}&G_{L}W\\G_{R}W^{\dagger}&\mathbf{1}+G_{R}\vec{h}\cdot\vec{\sigma}\tau_{3}
    \end{bmatrix}\Bigg) \\&= \text{Tr ln}\Bigg(\begin{bmatrix}
        \mathbf{1}&0\\G_{R}W^{\dagger}(\mathbf{1}+G_{L}\vec{h}\cdot\vec{\sigma}\tau_{3})^{-1}&\mathbf{1}
    \end{bmatrix}\begin{bmatrix}
        \mathbf{1}+G_{L}\vec{h}\cdot\vec{\sigma}\tau_{3}&0\\0&\mathbf{1}+G_{R}\vec{h}\cdot\vec{\sigma}\tau_{3}-G_{R}W^{\dagger}(\mathbf{1}+G_{L}\vec{h}\cdot\vec{\sigma}\tau_{3})^{-1}G_{L}W
    \end{bmatrix}\nonumber\\&\begin{bmatrix}
        \mathbf{1}&(\mathbf{1}+G_{L}\vec{h}\cdot\vec{\sigma}\tau_{3})^{-1}G_{L}W\\0&\mathbf{1}
    \end{bmatrix}\Bigg)\label{eq:A2}\\
    &= \text{Tr ln}\Bigg(\begin{bmatrix}
        \mathbf{1}&G_{L}W(\mathbf{1}+G_{R}\vec{h}\cdot\vec{\sigma}\tau_{3})^{-1}\\0&\mathbf{1}
    \end{bmatrix}\begin{bmatrix}
        \mathbf{1}+G_{L}\vec{h}\cdot\vec{\sigma}\tau_{3}-G_{L}W(\mathbf{1}+G_{R}\vec{h}\cdot\vec{\sigma}\tau_{3})^{-1}G_{R}W^{\dagger}&0\\0&\mathbf{1}+G_{R}\vec{h}\cdot\vec{\sigma}\tau_{3}
    \end{bmatrix}\nonumber\\&\begin{bmatrix}
        \mathbf{1}&0\\(\mathbf{1}+G_{R}\vec{h}\cdot\vec{\sigma}\tau_{3})^{-1}G_{R}W^{\dagger}&\mathbf{1}
    \end{bmatrix}\Bigg)\label{eq:A3}\;.
\end{align}
From this we derive based on Eq.~(\ref{eq:A2}),
\begin{align}
    S_T & = \text{Tr ln}\Big(\begin{bmatrix}
        \mathbf{1}+G_{L}\vec{h}\cdot\vec{\sigma}\tau_{3}&G_{L}W\\G_{R}W^{\dagger}&\mathbf{1}+G_{R}\vec{h}\cdot\vec{\sigma}\tau_{3}
    \end{bmatrix} = \text{ln det}\begin{bmatrix}
        \mathbf{1}+G_{L}\vec{h}\cdot\vec{\sigma}\tau_{3}&G_{L}W\\G_{R}W^{\dagger}&\mathbf{1}+G_{R}\vec{h}\cdot\vec{\sigma}\tau_{3}
    \end{bmatrix}\nonumber\\& = \text{ln det}\Big((\mathbf{1}+G_{L}\vec{h}\cdot\vec{\sigma}\tau_{3})(\mathbf{1}+G_{R}\vec{h}\cdot\vec{\sigma}\tau_{3})-(\mathbf{1}+G_{L}\vec{h}\cdot\vec{\sigma}\tau_{3})G_{R}W^{\dagger}(\mathbf{1}+G_{L}\vec{h}\cdot\vec{\sigma}\tau_{3})^{-1}G_{L}W\Big)\nonumber\\& = \text{Tr ln}\Big((\mathbf{1}+G_{L}\vec{h}\cdot\vec{\sigma}\tau_{3})(\mathbf{1}+G_{R}\vec{h}\cdot\vec{\sigma}\tau_{3})-(\mathbf{1}+G_{L}\vec{h}\cdot\vec{\sigma}\tau_{3})G_{R}W^{\dagger}(\mathbf{1}+G_{L}\vec{h}\cdot\vec{\sigma}\tau_{3})^{-1}G_{L}W\Big)\;,
\end{align}
and similarly, via Eq.~(\ref{eq:A3}),
\begin{align}
   S_T & = \text{Tr ln}\begin{bmatrix}
        \mathbf{1}+G_{L}\vec{h}\cdot\vec{\sigma}\tau_{3}&G_{L}W\\G_{R}W^{\dagger}&\mathbf{1}+G_{R}\vec{h}\cdot\vec{\sigma}\tau_{3}
    \end{bmatrix} \nonumber\\& = \text{Tr ln}\Big((\mathbf{1}+G_{R}\vec{h}\cdot\vec{\sigma}\tau_{3})(\mathbf{1}+G_{L}\vec{h}\cdot\vec{\sigma}\tau_{3})-(\mathbf{1}+G_{R}\vec{h}\cdot\vec{\sigma}\tau_{3})G_{L}W(\mathbf{1}+G_{R}\vec{h}\cdot\vec{\sigma}\tau_{3})^{-1}G_{R}W^{\dagger}\Big)\;.
\end{align}

Next, we can do, as in \cite{kamenev2023field} for the spin-independent case, the substitution $G_{L,R}\xrightarrow{}i\pi\nu Q_{L,R}$ for each channel. To obtain variation with $Q_{R}$, in the first expression take ($\mathbf{1}+G_{L}\vec{h}\cdot\vec{\sigma}$) out of the logarithm using the sum rule. Since the expression within brackets is linear in $Q_{R}$, the variation over the logarithm can be done.

Indeed, first of all notice that because of the product rule $\text{Tr ln}(AB) = \text{Tr ln}(A)+\text{Tr ln}(B)$, we may extract a factor $(1+Q_{L}\vec{h}\cdot\sigma\tau_{3})$ from the logarithm. Since this term does not depend on $Q_{R}$, we may omit it during the variation over $Q_{R}$. The resulting expression over which we vary has the form \begin{align}\sum_n\text{Tr ln}(\mathbf{1}+Q_{R}X_n)\;,\end{align} where \begin{align}X_n = i\pi\nu\vec{h}\cdot\sigma\tau_3+(\pi\nu)^{2}W_n^{\dagger}(\mathbf{1}+i\pi\nu Q_{L}\vec{h}\cdot\vec{\sigma}\tau_{3})^{-1}Q_{L}W_n\;.\end{align}

The inverse can not be calculated explicitly in general.

\begin{comment} 
The inverse can be computed explicitly for the expression under considerations, due to  the two identities $Q_{L}^{2} = \mathbf{1}$ and $(\vec{h}\cdot\vec{\sigma}) = h^{2}\mathbf{1}$, and equals $\frac{1}{1+h^{2}}(1-i\vec{h}\cdot\sigma\tau_{3}Q_{L})$. From this expression we conclude that the inverse always exists and consequently the procedure above can always be followed. 

With this,
\begin{align}
    X = i\pi\nu\vec{h}\cdot\vec{\sigma}\tau_{3}+(\pi\nu)^{2}W^{\dagger}Q_{L}W-\frac{(\pi\nu)^{2}}{1+h^{2}}W^{\dagger}i\pi\nu\vec{h}\cdot\vec{\sigma}\tau_{3}W.
\end{align}
\end{comment}

Therefore, variation of $Q_{R}$ via $Q_{R}\xrightarrow{}Q_{R}+[Q_{R},\delta Q]$ gives, up to first order in $\delta Q$
\begin{align}
    S_T(Q_{R}+[Q_{R},\delta Q])-S_T(Q_R)&=\text{Tr}\Bigg(-\sum_{k = 1}^{\infty}\frac{(-1)^{k}}{k}(((Q_R+[Q_{R},\delta Q])X_{n})^{k}-(Q_R X_{n})^k) \Bigg)\nonumber\\&= \text{Tr}\Bigg(-\sum_{k = 1}^{\infty}(-1)^{k} (X_{n}Q_{R})^{k}\delta Q-(Q_{R}X_{n})^{k}\delta Q\Bigg)\nonumber\\& = \text{Tr}\Bigg(\sum_{k = 1}^{\infty}[(-X_{n}Q_{R})^{k-1}X_{n},Q_{R}] \delta Q\Bigg)\nonumber\\&= \text{Tr}\Bigg([(1+X_{n}Q_{R})^{-1}X_{n},Q_{R}]\delta Q\Bigg)\;.
\end{align}
where the factor $k$ disappears from the first to second line because there are $k$ positions where we can insert $\delta Q$.
Since the matrix current is defined such that for all $\delta Q$ we have
\begin{align}
    S_T(Q_{R}+[Q_{R},\delta Q])-S_T(Q_R) = \text{Tr}(I_R\delta Q)\;,
\end{align}
we conclude that
\begin{align}
    I_{R} &=[(1+X_{n}Q_{R})^{-1}X_{n},Q_{R}]\\
    &=[(1+\Big(i\pi\nu\vec{h}\cdot\vec{\sigma}\tau_3+(\pi\nu)^{2}W_{n}^{\dagger}Q_{L}W_{n}-i(\pi\nu)^{3} W_{n}^{\dagger}\vec{h}\cdot\vec{\sigma}\tau_{3}W_{n}\Big)Q_{R})^{-1}\Big(i\pi\nu\vec{h}\cdot\sigma\tau_3-(\pi\nu)^{2}W_{n}^{\dagger}(\mathbf{1}+i\pi\nu Q_{L}\vec{h}\cdot\vec{\sigma}\tau_{3})Q_{L}W_{n}\Big),Q_{R}]\;.
\end{align}

This corresponds to Eq.~(\ref{eq:magn_X}) in the main text.
\section{Obtaining the surface Green's function}\label{sec:CalculatingC}
In this Appendix, we show how to obtain the surface Green's function $C$ in Eq.~(\ref{eq:Cexpression}) for a homogeneous material on the left hand side.

As indicated in the main text, Eqs.~(\ref{eq:gLpbulk},~\ref{eq:gLmbulk}) the Eilenberger equations for the incoming and outgoing modes reads
\begin{align}
    v_F\cos\theta\partial_x g_{L+} = \alpha_1 [\Lambda_1,g_{L+}]\;,\label{eq:gLpc}\\
    v_F\cos\theta\partial_x g_{L-} = -\alpha_2 [\Lambda_2,g_{L-}]\label{eq:glmc}\;,
\end{align}
where $\Lambda_{1,2}$ are spatially independent matrices whose square is proportional to the identity matrix, $v_F$ is the Fermi velocity and $\alpha_{1,2}$ are energy scales determined by the intrinsic energy scales of the material.
while they obey the boundary condition
\begin{align}
    g_{L+}(x = 0) = g_{L-}(x = 0) =:C\;. \label{eq:bcgLpgLm}
\end{align}
Moreover, the elements of $g_{\pm}$ must not diverge as $x\xrightarrow{}-\infty$, but instead converge to the elements of $\Lambda_{1,2}$.

We may always use a transformation to a basis in which $\Tilde{\Lambda}_1 = \tau_3$, where we use $\tilde{\cdot}$ to distinguish matrices using the alternative basis. Moreover, in this same basis we denote $\tilde{C} = \begin{bmatrix}
    \tilde{C}_{11}&\tilde{C}_{12}\\\tilde{C}_{21}&\tilde{C}_{22}
\end{bmatrix}$, where  $\tilde{C}_{ij}$ are matrices in spin space. Within this basis, any solution that obeys Eq.~(\ref{eq:gLpc}) and Eq.~(\ref{eq:bcgLpgLm}) satisfies
\begin{align}
    \tilde{g}_{L+}(x) = \begin{bmatrix}
        1 + (\tilde{C}_{11}-1)&\tilde{C}_{12}e^{2\alpha_1 x /(\cos\theta)}\\\tilde{C}_{21}e^{-2\alpha_1 x/(\cos\theta)}&-1 +(\tilde{C}_{22}+1)
    \end{bmatrix}\;.\label{eq:gLtildebasis}
\end{align}
Here $\theta$ is the angle between the incoming mode and the normal to the interface.
Requiring that $\tilde{g}_{L+}(x)$ converges to $\tilde{\Lambda}_1$ $x\xrightarrow{}-\infty$ requires (among others) $\tilde{C}_{21} = 0$ and $\tilde{C}_{22} + 1 = 0$, which can be written as $(1-\tilde{\Lambda}_1)(\tilde{C} + 1) = 0$. In the original basis this condition reads \begin{align} (1-\Lambda_1)(C + 1) = 0\;,\label{eq:Ccond1}\end{align} which produces Eq.~(\ref{eq:Cc1}) in the main text.

Entirely similar, we may change a basis, denoted using $\bar{\cdot}$ in which $\bar{\Lambda}_2 = \tau_3$. In this basis
\begin{align}
    g_{L-}(x) = \begin{bmatrix}
        1 + (\bar{C}_{11} - 1)& \bar{C}_{12}e^{-2\alpha_2 x/(\cos\theta)}\\\bar{C}_{21}e^{2\alpha_1 x/(\cos\theta)} & -1 + (\bar{C}_{22} + 1)
    \end{bmatrix}\;.\label{eq:glbarbasis}
\end{align}
This converges only to $\bar{\Lambda}_2$ if (amongst others) $\bar{C}_{11}-1 = 0$ and $\bar{C}_{12} = 0$. This can be written as $(1+\bar{\Lambda}_2)(\bar{C} -1)$, which implies
\begin{align} (1+\Lambda_2)(C - 1) = 0\;.\label{eq:Ccond2}\end{align} This is Eq.~(\ref{eq:Cc2}) in the main text.
From Eqs.~(\ref{eq:Ccond1},~\ref{eq:Ccond2}), it follows that
\begin{align}
    (1-\Lambda_{1})C &= -1+\Lambda_{1}\;,\label{eq:labmda1eq}\\
    (1+\Lambda_{2})C &= 1+\Lambda_{2}\;,\label{eq:lambda2eq}
\end{align}
and hence via subtraction of these two equations
\begin{align}
    (\Lambda_{2}+\Lambda_{1})C = 2+\Lambda_{2}-\Lambda_{1}\;.
\end{align}
Defining $H_{\pm} = \frac{1}{2}(\Lambda_1\pm\Lambda_2)$, this reads
\begin{align}
    C = H_{+}^{-1}(1-H_-)\label{eq:Cexpressionapp}\;,
\end{align}
which is Eq.~(\ref{eq:Cexpression}) in the main text.

We conclude that $C$ is uniquely determined by just a subset of the constraints following from Eqs. (\ref{eq:gLtildebasis},~\ref{eq:glbarbasis}). In what follows, we show that this expression for $C$ is also consistent with the remaining constraints. 

The first additional constraint is obtained by addition of Eqs. (\ref{eq:labmda1eq},~\ref{eq:lambda2eq}), which gives
\begin{align}
    C = (1-H_{-})^{-1}H_{+} = \Big(H_{+}^{-1}(1-H_{-})\Big)^{-1} = C^{-1}\;.
\end{align}
%As derived in \cite{tanaka2022theory} this satisfies $C^2 = \mathbf{1}$.
Thus, this constraint is satisfied as long as $C^2 = \mathbf{1}$. As shown in \cite{tanaka2022theory}, $C$ satisfies this property.

Thus, the constructed $C$ satisfies the constraints $\tilde{C}_{21} = 0$, $\tilde{C}_{22}+1 = 0$, $\bar{C}_{12} = 0$ and $\bar{C}_{11}-1 = 0$.
However, following Eqs.~(\ref{eq:gLtildebasis},~\ref{eq:glbarbasis}), the surface Green's function additionally has to satisfy the  constraints $\tilde{C}_{11} - 1 = 0$ and $\bar{C}_{22} + 1$ = 0.  Therefore, combining $\tilde{C}_{11}-1 = 0$ with $\tilde{C}_{21} = 0$, we find that in analogy to the derivation above that $C$ has to satisfy
\begin{align}
    (C-1)(1+\Lambda_1) = 0\;.\label{eq:lambda1eq2}
\end{align}
Similarly, exploiting $\bar{C}_{22} = 0$ and $\bar{C}_{12} = 0$, 
\begin{align}
    (C+1)(1-\Lambda_{2}) = 0\;.\label{eq:labmda2eq2}
\end{align}

Addition and subtraction of Eqs.~(\ref{eq:lambda1eq2},~\ref{eq:labmda2eq2}) give, exploiting $\{H_{+},H_{-}\} = 0$,
\begin{align}
    C &= H_{+}(1+H_{-})^{-1} = (1-H_{-})^{-1}H_{+} = \Big(H_{+}^{-1}(1-H_{-})\Big)^{-1} = H_{+}^{-1}(1-H_{-})\;,\\
    C &= (1+H_{-})H_{+}^{-1} = H_{+}^{-1}(1-H_{-})\;.
\end{align}

In conclusion, all of the constraints on $C$ are satisfied if and only if
\begin{align}
    C = H_{+}^{-1}(1-H_{-})\;.
\end{align}
which is Eq.~(\ref{eq:Cexpression}) in the main text.

%\includepdf{Pwave_SpinSplitJunction_Overleaf.pdf}
\section{Surface Green's function with interfacial exchange field}\label{sec:XfromCexchangefield}
In this Appendix we derive Eqs. (\ref{eq:Xperp}) and (\ref{eq:Xparallel}) for the surface Green's function modified by the interfacial exchanged field.

We start from Eq.~(\ref{eq:XC}) in the main text, 
\begin{align}
    X = i\vec{h}\cdot\vec{\sigma}\tau_3 + T_{1}(1+i\pi\nu C\vec{h}\cdot\vec{\sigma}\tau_3))^{-1}C = i\vec{h}\cdot\vec{\sigma}\tau_3+T_{1} (C+i\pi\nu \vec{h}\cdot\vec{\sigma}\tau_3))^{-1}\;.
\end{align}

We choose our axes in spin space such that $\vec{h} = h\vec{z}$. We consider separately the cases in which $\vec{d}$ is parallel to $\vec{z}$ and perpendicular to $\vec{z}$.

\subsection{Perpendicular d-vector}
Suppose $\vec{d}$ lies in the $xy$-plane, $\vec{d} = (\cos\chi,\sin\chi,0)$. In this case, using the basis $(\psi_{\uparrow},\psi_{\downarrow},\psi^{\dagger}_{\downarrow},-\psi^{\dagger}_{\uparrow})$, we have
\begin{align}
    C = \frac{\Delta}{E}\tau_3 + \frac{i\sqrt{\Delta^2-E^2}}{E}\tau_1(\cos\chi \sigma_x + \sin\chi \sigma_y)\;.
\end{align}
Thus, $C$ commutes with $\vec{h}\cdot\vec{\sigma}\tau_3$, so that
\begin{align}
    (\vec{C}+i\vec{h}\cdot\vec{\sigma}\tau_3)(\vec{C}-i\vec{h}\cdot\vec{\sigma}\tau_3) = \vec{C}^2 + (\vec{h}\cdot\vec{\sigma}\tau_3)^2 = (1+(\pi\nu h)^2)\mathbf{1}\;.
\end{align}
Therefore, we may write
\begin{align}
    X = i\pi\nu\vec{h}\cdot\vec{\sigma}\tau_3 + \frac{T_{1}}{1+(\pi\nu h)^2}(C-i\pi\nu\vec{h}\cdot\vec{\sigma}\tau_3)\;.
\end{align}
For small energies the elements of $C$ become large, so that,
\begin{align}
    X\approx \frac{T_{1}}{1+(\pi\nu h)^2}C\;.
\end{align}
With this, we derive Eq.~(\ref{eq:Xperp}) in the main text.
\subsection{Parallel d-vector}
Now suppose $\vec{d}$ is parallel to $\vec{h}$, that is
\begin{align}
    \vec{C} = \frac{\Delta}{E}\tau_3 + \frac{i\sqrt{\Delta^2-E^2}}{E}\tau_1\sigma_z\;.
\end{align}
In this case, both $C$ and $\vec{h}\cdot\vec{\sigma}$ commute with $\sigma_z$, so that we may decompose into two by two blocks:
\begin{align}
    C + i\pi\nu h\sigma_z\tau_3 = \frac{1+\sigma_z}{2}\otimes\Big((\frac{\Delta}{E}+i\pi\nu h)\tau_3 + \frac{i\sqrt{\Delta^2-E^2}}{E}\tau_1\Big) + \frac{1-\sigma_z}{2}\otimes\Big((\frac{\Delta}{E}-i\pi\nu h)\tau_3 - \frac{i\sqrt{\Delta^2-E^2}}{E}\tau_1\Big)\;.
\end{align}
where the symbol $\otimes$ denotes the Kronecker product.

Written in this form, we may directly apply the inversion rules for two by two matrices and find
\begin{align}
    (C+i\pi\nu h\sigma_z\tau_3)^{-1} &= \frac{1+\sigma_z}{2}\otimes\frac{1}{1 + \frac{2i\pi\nu h\Delta}{E}-(\pi\nu h)^2}\Big((\frac{\Delta}{E}+i\pi\nu h)\tau_3 + \frac{i\sqrt{\Delta^2-E^2}}{E}\tau_1\Big) \nonumber\\&+ \frac{1-\sigma_z}{2}\otimes\frac{1}{1 - \frac{2i\pi\nu h\Delta}{E}-(\pi\nu h)^2}\Big((\frac{\Delta}{E}-i\pi\nu h)\tau_3 - \frac{i\sqrt{\Delta^2-E^2}}{E}\tau_1\Big)\;.
\end{align}
For low energies, this reduces to, keeping only terms to lowest order in $E$, 
\begin{align}
    (C+i\pi\nu h\sigma_z\tau_3)^{-1} \approx \frac{1+\sigma_z}{2}\otimes\Big(-\frac{i}{2\pi\nu h}(\tau_3 + i\tau_1) \Big)+\frac{1-\sigma_z}{2}\otimes\Big(\frac{i}{2\pi\nu h}(\tau_3 - i\tau_1) \Big) = \frac{-i}{2\pi\nu h}\vec{\hat{h}}\cdot\vec{\sigma}\tau_3 + \frac{1}{2\pi\nu h}\tau_1\;,
\end{align}
Plugging this into $X$, we obtain
\begin{align}
    X = i\pi\nu(1-\frac{T_{1}}{2(\pi\nu h)^2})\vec{h}\cdot\vec{\sigma}\tau_3 + \frac{T_{1}}{2\pi\nu h}\tau_1\;,
\end{align}
which is Eq.~(\ref{eq:Xparallel}) in the main text.

\end{document}